\documentclass[amsmath,amssymb,11pt]{article}
\usepackage{jheppub2}
\pdfoutput=1
\usepackage{graphicx,epsfig}
\usepackage{float}
\usepackage{subfig}
\usepackage{subfloat}
\usepackage[utf8]{inputenc}
\usepackage{hyperref}
\usepackage{caption}
\usepackage{color}
\usepackage{overpic}
\usepackage[dvipsnames]{xcolor}
\usepackage{physics}
\usepackage[multiple]{footmisc}

\usepackage{mathtools}

\newcommand{\nocontentsline}[3]{}
\newcommand{\tocless}[2]{\bgroup\let\addcontentsline=\nocontentsline#1{#2}\egroup}

\def\ie{{\it i.e.} }
\def\eg{{\it e.g.} }

\newcommand{\df}{\mathrm{d}}

\newcommand{\beq}{\begin{equation}}
\newcommand{\eeq}{\end{equation}}

\newcommand{\be}{\begin{equation}}
\newcommand{\ee}{\end{equation}}

\usepackage{tikz}
\usetikzlibrary{positioning}
\usetikzlibrary{intersections}
\usetikzlibrary{fadings} 
\usetikzlibrary{arrows.meta} 
\usetikzlibrary{arrows}

\tikzfading[name=fade out,
inner color=transparent!0,
outer color=transparent!100]

\definecolor{cherryblossompink}{rgb}{1.0, 0.72, 0.77}
\definecolor{lightblue}{rgb}{0.68, 0.85, 0.9}

\usetikzlibrary{decorations.pathmorphing}
\usetikzlibrary{decorations.pathreplacing,decorations.markings}

\usetikzlibrary{backgrounds,automata}

\title{Gravity in Enhanced Brane-World Models}

\author{Quim Llorens}
\emailAdd{joaquim.llorens-giralt@uni-wuerzburg.de}

\affiliation{
Institute for Theoretical Physics and Astrophysics, \\
Julius Maximilians Universit\"at W\"urzburg, Am Hubland, 97074 W\"urzburg
}

\abstract{
We generalize and extend results on the localization of gravity on Karch-Randall-Sundrum brane-worlds with positive, negative, or zero cosmological constant on the brane. We do so both from the study of bulk metric perturbations, and from their reinterpretation through brane-world holography ---an induced higher-derivative theory of gravity coupled to a cut-off CFT on the brane.
We then enhance these models by adding an explicit Einstein-Hilbert term on the brane action ---a DGP term--- and, through studying the brane position and the localization of gravity on the brane, we establish bounds for its coupling constant, beyond which the theory presents pathologies. We finally study the limit in which the brane reaches the boundary, and comment on adding further higher-derivative terms on the brane action.
}

\begin{document}

\maketitle

%%%%%%%%%%%%%%%%%%%%%%%%%%%%%%%%%%%%%%%%%%%%%%%%%%%%%%%
%
%
%
%
%
%%%%%%%%%%%%%%%%%%%%%%%%%%%%%%%%%%%%%%%%%%%%%%%%%%%%%%%

\section{Introduction}

%%%%%%%%%%%%%

To explain our four-dimensional world with superstring theory, six spatial dimensions need to be hidden away.
Besides compactification, gravity can be localized on a brane either thanks to the curvature of the bulk, or by adding an explicit Einstein-Hilbert term on the brane.

Randall and Sundrum (RS) \cite{Randall:1999vf} proposed a four-dimensional purely-tensional flat brane in a five-dimensional Anti-de Sitter (AdS) bulk. Thanks to the bulk curvature, a bulk graviton mode localizes on the brane.
This was soon generalized to AdS and dS branes \cite{Karch:2000ct} by Karch and Randall (KR).
Dvali, Gabadadze, and Porrati (DGP) \cite{Dvali:2000hr} introduced a different mechanism, embedding a four-dimensional flat brane with an explicit Einstein-Hilbert term in its action into a five-dimensional Minkowski bulk.
The KRS models reproduce four-dimensional gravity at long distances, while the DGP model does the opposite \cite{Garriga:1999yh, Tanaka:2003zb}.
In this work, we will study DGP branes sitting on an AdS bulk, as an extension of the KRS framework \cite{Randall:1999vf, Karch:2000ct}.

The holographic duality \cite{Maldacena:1997re, Gubser:1999vj, Witten} can also be applied to KRS brane-worlds \cite{Verlinde:1999fy, Hawking:2000kj, Porrati:2001gx, deHaro:2000wj}, in what is known as brane-world holography: the AdS$_{d+1}$ bulk ending on an end-of-the-world (EOW) brane is dual to an effective $d$-dimensional gravitational theory coupled to a CFT$_d$ on the brane \cite{Neuenfeld:2021wbl}.
Since the brane is at a finite distance from the boundary, the dual CFT has a UV cut-off.
Moreover, the induced gravity theory on the brane is actually a higher-derivative theory of gravity \cite{deHaro:2000wj, Neuenfeld:2021wbl, Bueno:2022log}, with the structure of the higher-derivative terms being fixed by the bulk AdS asymptotics ---in fact, they take the same form as the counterterms for holographic renormalization in standard AdS/CFT \cite{deHaro:2000vlm, Emparan:1999pm, Kraus:1999di, Skenderis:2002wp, Bueno:2022log}.
We will reinterpret the localization of gravity on the brane from this holographic perspective, both with and without a DGP term. 

In this work, we aim to better understand DGP-KRS brane-worlds and their range of applicability.
Already \cite{Tanaka:2003zb} studied DGP-KRS brane-worlds soon after the original works, but in the context of cosmology, not holography.
More recently, \cite{Geng:2023iqd, Perez-Pardavila:2023rdz, Geng:2023qwm} constrained the allowed values for the DGP coupling through studying the entanglement entropy of brane subregions.
Similarly, \cite{Lee:2022efh} studied the information-theoretic properties of two-dimensional brane-worlds in planar black hole spacetimes, and put bounds on the allowed values of the couplings on the brane.
Finally, in \cite{Miao:2023mui}, the author placed constraints on the DGP coupling (and curvature squared couplings) through studying the inner product of the bulk graviton modes.

We will obtain similar results to \cite{Geng:2023iqd} and \cite{Miao:2023mui}: the DGP coupling cannot be too negative ---in our conventions (see Section \ref{chp:BWsWithDGP}), the DGP coupling $A$ must fulfil $A > -1/2$.
We will obtain this results in two separate ways. 
Firstly, by studying the position of the brane as a function of the DGP coupling $A$, while keeping the brane tension fixed.
And secondly, by explicitly calculating, analytically and numerically, the mass of the graviton modes on AdS branes as a function of the DGP coupling and the brane position.

Another motivation behind this work is making contact with models of dynamical-gravity holography \cite{Compere:2008us, Ishibashi:2023luz, Ishibashi:2023psu, Ishibashi:2024fnm, Ghosh:2023gvc, Ecker:2021cvz, Ecker:2023uea, Parvizi:2025shq, Parvizi:2025wsg}.
These models contain no brane but include an explicit Einstein-Hilbert term on the boundary action, making the boundary metric dynamical under mixed boundary conditions.
We want to study these models as the limit of DGP-KRS brane-worlds in which the brane is sent to the boundary.
However, the physical metric on the brane diverges at the boundary and must be rescaled, and we must also add counterterms to regulate the CFT partition function.
In particular, we must add a counterterm which can be reinterpreted as a DGP term with precisely $A = -1/2$ at the regulating hypersurface.
While this observation does not directly invalidate these dynamical-boundary models ---it tells us that the extra Einstein-Hilbert term on the boundary must have a positive coupling, which is something that is already known--- it makes the comparison between both models less straightforward.

\paragraph*{Summary.} 
We start by reviewing the localization of gravity on Karch-Randall-Sundrum brane-worlds, in which a $d$-dimensional brane sits near the boundary of an AdS$_{d+1}$ bulk. 
We reformulate and generalize the KRS model to deal simultaneously with all three maximally symmetric brane geometries and any number of dimensions $d \geq 3$. 
For AdS branes, we improve the formula of the brane graviton mass and present new results on the mass of the higher overtones.
We then reinterpret our findings through brane-world holography.
While we do not deviate strongly from \cite{Karch:2000ct, Randall:1999vf} and their holographic reformulations \cite{deHaro:2000vlm, Neuenfeld:2021wbl}, we generalize and simplify the computations while clarifying their interpretation.

We then add a DGP term ---an explicit Einstein-Hilbert term--- on the brane action. 
By studying how it changes the localization of gravity on the brane, we establish bounds for the DGP coupling constant:
a positive coupling strengthens gravity localization, while a negative coupling weakens gravity on the brane up to a point beyond where the construction breaks down ---either the position of the brane ceases to be well-defined, if we fix the brane tension; or its linearized spectrum presents a negative-mass mode, if we fix the brane position.

Next, we study the limit in which the brane is sent to the boundary. 
While our results here are inconclusive, they suggest that DGP-KRS brane-worlds may not be useful to interpolate between brane-world holography and dynamical-boundary set-ups ---when doing brane-world holography, we crucially do not add counterterms to our action, but they are needed to render the CFT action finite in dynamical-boundary models, and their coupling values exactly lie on the problematic point beyond which we can no longer trust DGP-KRS models.

We finally comment on the addition of further higher-curvature corrections on the brane.

%%%%%%%%%%%%%%%%%%%%%%%%%%%%%%%%%%%%%%%%%%%%%%%%%%%%%%%%%%%%%%%%%%%%%%%%%%%%%%%%%%%%%%%%%%%%%
%
%
%
%
%
%
%
%
%
%%%%%%%%%%%%%%%%%%%%%%%%%%%%%%%%%%%%%%%%%%%%%%%%%%%%%%%%%%%%%%%%%%%%%%%%%%%%%%%%%%%%%%%%%%%%

\begin{figure}[t]
    \centering
    \includegraphics[scale=0.42]{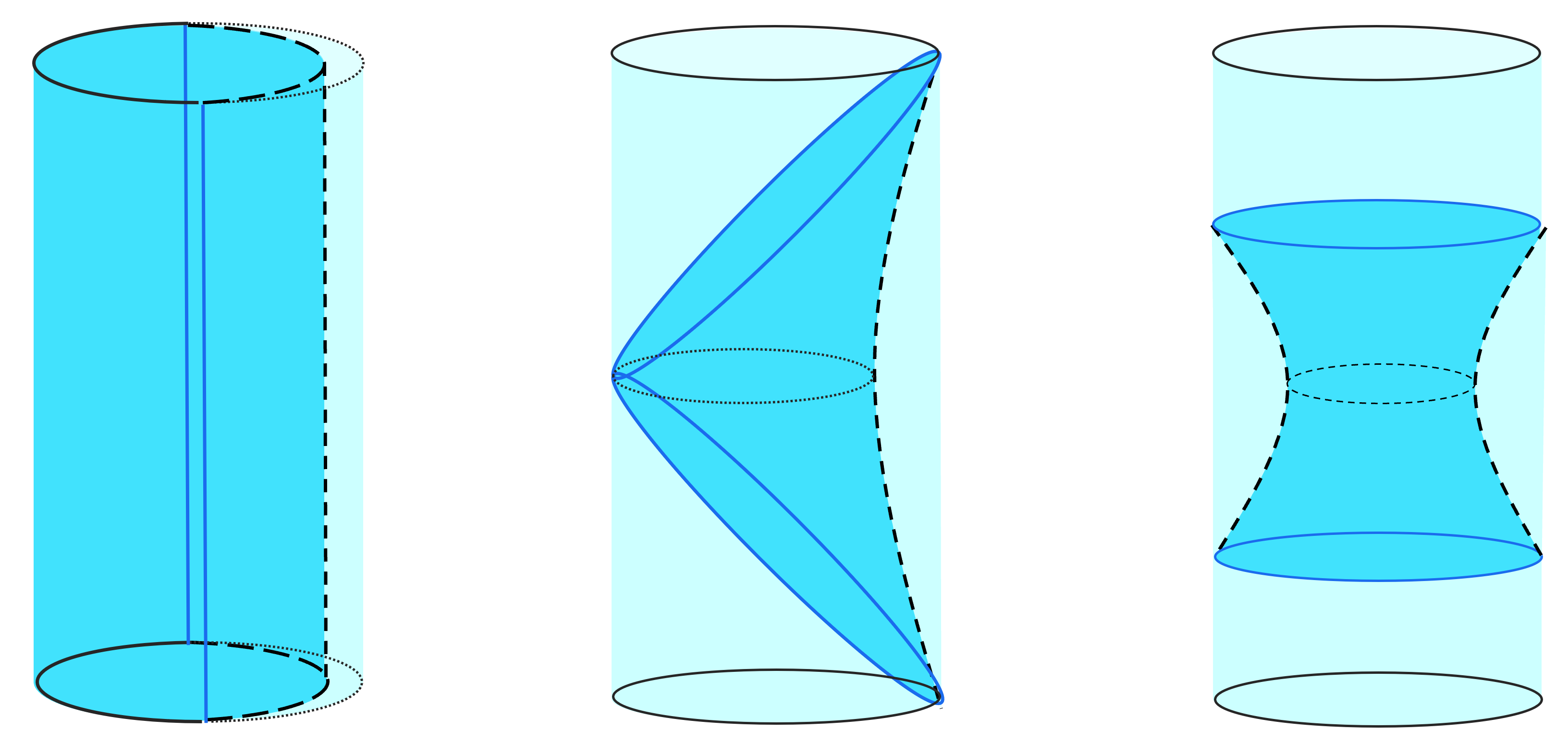}
    \caption{\small{The three maximally symmetric brane-worlds in global AdS. The part of the bulk spacetime in light blue is removed from the set-up. 
    Left: AdS brane. Center: Flat brane. Right: dS brane.
    }}
    \label{fig:Branes}
\end{figure}

%%%

\section{Gravity on KRS Brane-Worlds}
\label{sec:IntroBWs}

The Randall-Sundrum model \cite{Randall:1999vf} consists in embedding a four-dimensional flat brane into a five-dimensional AdS spacetime.
Due to the bulk curvature, a massless graviton mode localizes on the brane. 
Although there is also a continuum of Kaluza-Klein modes, their wave-function is suppressed on the brane.
Thus, gravity on the brane is effectively four-dimensional at low energies, displaying a Newtonian potential at large scales \cite{Garriga:1999yh}.
However, at high energies, \ie at distances smaller than the bulk AdS radius, gravity remains higher-dimensional.

To get a flat brane geometry, Randall and Sundrum had to fine-tune its tension.
Soon afterwards, Karch and Randall \cite{Karch:2000ct} showed that gravity still localizes on the brane even if this requirement is relaxed.
A subcritical brane tension leads to an AdS brane geometry, while a supercritical tension results in a dS brane.
Nevertheless, if the tension is close enough to criticality, then the brane sits near the asymptotic boundary and one recovers almost-flat four-dimensional gravity strongly localized on the brane.

Although there exist some top-down brane-world set-ups (see  \eg \cite{Verlinde:1999fy, Chiodaroli:2012vc, Gutperle:2020gez, Uhlemann:2021nhu, Karch:2022rvr}), most brane-world constructions are bottom-up models in which the brane is infinitely thin and purely tensional, with no other charges.
From an EFT perspective, the brane tension is only the first term of a series expansion describing the brane action, and so we could and should consider adding higher-order operators to it, as we will do in following sections.

One can reinterpret brane-world models through holography, as an effective gravity theory in four dimensions coupled to cut-off CFT radiation dual to the five-dimensional AdS bulk \cite{Porrati:2001gx, deHaro:2000wj}.
Again, this effective brane picture is clearest when the brane is close to the boundary, since the CFT cut-off is related to the renormalized distance from the brane to the boundary. 
The holographically-induced theory on the brane is not simply Einstein gravity, but a higher-derivative theory \cite{deHaro:2000wj, Neuenfeld:2021wbl}. 
The terms in the expansion can be computed algorithmically from the bulk \cite{Bueno:2022log}, following the same procedure used in finding the counterterms for holographic renormalization \cite{deHaro:2000wj, Kraus:1999di, Skenderis:2002wp, Emparan:1999pm}.
Again, the parameter controlling the higher-derivative expansion in the effective action is related to the distance from the brane to the boundary.

Recent research on brane-worlds has focused on of AdS branes, which are qualitatively different from their flat or dS counterparts. 
This is mainly due to the brane only cutting off part of the boundary, as opposed to the other two cases, in which the boundary is completely removed. 
In dual terms, this means that AdS brane-worlds are not only dual to a cut-off CFT living on the dynamically gravitating brane geometry, but that this CFT is also coupled to a CFT living on the fixed geometry of the remaining asymptotic boundary, through transparent boundary conditions at the defect where the brane reaches the boundary.

A key feature of AdS brane-worlds is that the bulk graviton mode that localizes on the brane is massive \cite{Karch:2000ct, Miemiec:2000eq, Schwartz:2000ip}. 
From the bulk perspective, this graviton gets a mass because it is a mixture of a normalizable and a non-normalizable mode \cite{Neuenfeld:2021wbl},\footnote{Both modes are normalizable, since the brane sits at a finite distance from the boundary. However, if there were no brane, one of the modes would diverge, so we still use this distinction.} as opposed to the flat and dS brane cases, in which the brane graviton is massless since it comes from a purely non-normalizable bulk mode.
From the brane perspective, the graviton on the brane acquires a mass due to its interaction with the CFT radiation \cite{Porrati:2001gx}, which has transparent boundary conditions where the brane reaches the boundary.\footnote{
This is a general feature of gravity plus matter in AdS with transparent boundary conditions at infinity. If CFT radiation leaks out of the spacetime, the graviton gets a mass through a Higgs-like mechanism \cite{Porrati:2001db, Porrati:2003sa}.}
Furthermore, in the case of AdS brane-worlds, the Kaluza-Klein modes do not form a continuum as in the flat or dS cases, but a discrete spectrum \cite{Karch:2000ct}.
This is due to the Dirichlet boundary conditions at the asymptotic boundary, so bulk modes feel as if they were trapped in a potential well.

\begin{figure}[t]
    \centering
    \includegraphics[scale=0.46]{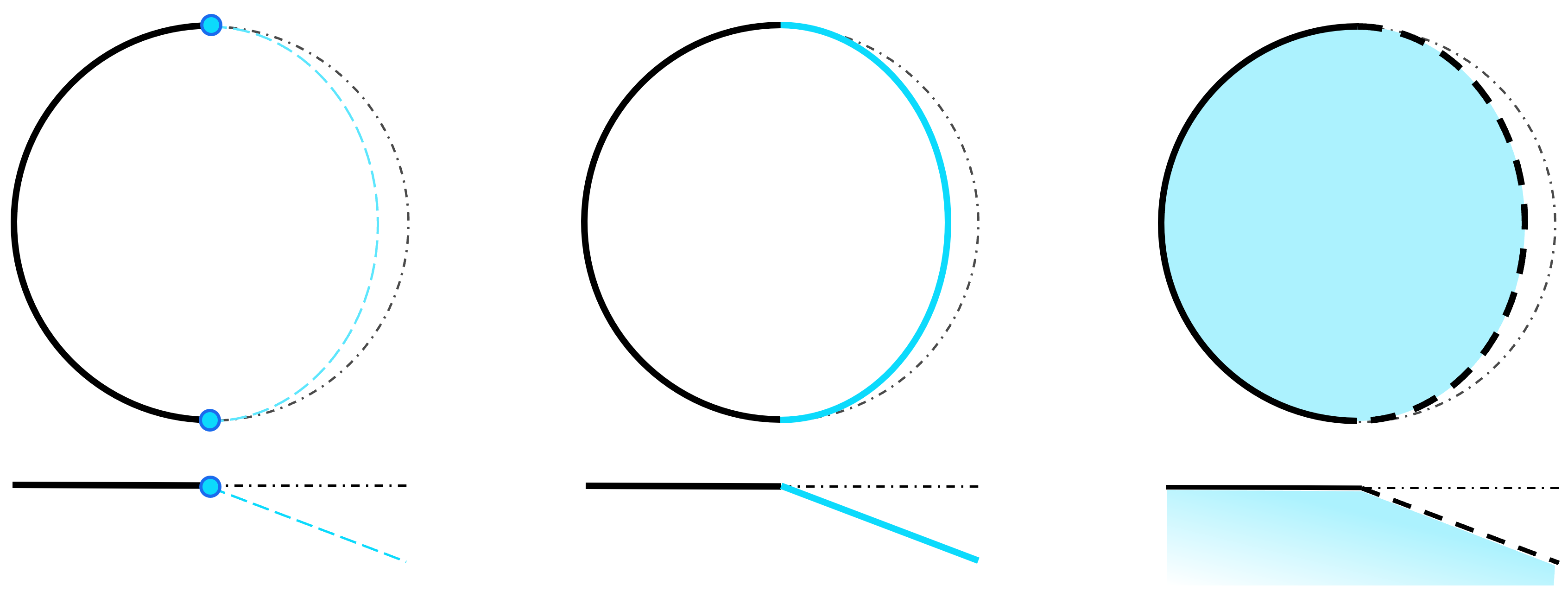}
    \caption{\small{The three different perspectives of double holography. 
    Top pictures are of a (global) time slice, while bottom ones are in Poincaré-like coordinates. 
    Left: Boundary perspective: BCFT$_d$, a CFT$_d$ living on a $d$-dimensional fixed spacetime (black) coupled to a CFT$_{d-1}$ at the boundary (blue dots).
    Middle: Brane perspective: CFT$_d$ living on the fixed geometry at the boundary (black) plus a cut-off CFT$_d$ coupled to (higher-derivative) gravity on the brane geometry (blue). There are transparent boundary conditions where the brane meets the boundary.
    Right: Bulk perspective: Einstein gravity in an AdS$_{d+1}$ spacetime (blue), containing an AdS$_d$ brane as an end-of-the-world brane (dotted line).
    }}
    \label{fig:KRPoinc}
\end{figure}

%%%%%%%%%

In this section, we will redo and expand the works of Randall, Sundrum, and Karch \cite{Randall:1999vf, Karch:2000ct}, simplifying the computations and generalizing them to any number of spacetime dimensions, while dealing simultaneously with all three maximally symmetric brane geometries. 
We will refine the formula of the graviton mass for AdS branes, and give new expressions for the mass of the higher harmonics. 
We will also discuss how to reinterpret these results from the brane perspective \cite{deHaro:2000wj, Neuenfeld:2021wbl}. 
Even though we will study all three brane cases, we will only give a detailed discussion of the AdS case, since it is the one that is more relevant for holography.

%%%%%%%%%%%%%%%%%%%%%%%%%%

\subsection{Set-up and Background Solution}\label{sec:SetUp}
Our starting point is the action
\begin{equation}\label{ActionBulk}
    I = I_\text{bulk} + I_\text{brane}\,,
\end{equation}
consisting of Einstein gravity in an AdS$_{(d+1)}$ bulk $\mathcal{M}$ ending on a brane $\partial\mathcal{M}_b$ with tension $\tau$,
\begin{align}\label{IBulk}
    I_{\text{bulk}} & = \frac{1}{16 \pi G_N} \left[ \int_{\mathcal{M}} \df^{d+1} x \sqrt{-G} \left(  R[G] - 2\Lambda \right) + 2 \int_{\partial\mathcal{M}} \df^dx \sqrt{-g} \ K \right]\,,\\
    I_\text{brane} & = - \int_{\partial\mathcal{M}_b} \df^dx \sqrt{-g} \ \tau\,,
    \label{IBrane}
\end{align}
where $G$ denotes the bulk metric, $g$ is the induced metric on $ \partial\mathcal{M}$, $G_N$ is the bulk Newton's constant, and $\Lambda$ is the bulk cosmological constant, with curvature radius $L$. 
The boundary of $\mathcal{M}$ is $\partial\mathcal{M} \supseteq \partial\mathcal{M}_b$, where we have included the Gibbons-Hawking-York (GHY) term explicitly to get the desired boundary conditions.
From now on, however, when we speak about the boundary, we will mean the asymptotic boundary at infinity, not the brane. 

%%%%%%

Varying \eqref{ActionBulk} with respect to $G_{MN}$ and imposing Dirichlet boundary conditions at infinity and Neumann boundary conditions on the brane, we get the usual bulk Einstein equations,
\begin{equation}\label{EEsBulk}
    R_{MN} [G] - \frac{1}{2} G_{MN} R[G] + \Lambda G_{MN} = 0\,,
\end{equation}
plus the Israel junction condition on the brane \cite{Israel:1966rt},
\begin{equation}\label{IJC}
    K_{ab} = \frac{8 \pi G_N \tau}{d-1} g_{ab}\,.
\end{equation}
where $K_{ab}$ is the extrinsic curvature on the brane.
This greatly restricts the geometry of the brane and its location ---the brane is forced to sit on a totally umbilic hypersurface.

Our ansatz for the bulk metric in Poincaré-like coordinates takes the form
\begin{equation}\label{SlicingMetric}
    d{s}^2_{d+1} = G_{\mu \nu} (x,z) dy^\mu dy^\nu = \frac{L^2}{\left(f(z)\right)^2} \left[dz^2 + \hat{g}_{ij}(x)dx^idx^j \right]\,,
\end{equation}
where the $d$-dimensional metric $\hat{g}_{ij}(x)$ is either flat, or an (A)dS$_d$ metric with unit radius.
This metric slices the bulk at constant $z$ into $d$-dimensional maximally symmetric slices. 
The function $f(z)$ is a function which behaves as $f(z) \sim z$ for small values of $z$,
\begin{equation}\label{f(z)Def}
    f(z) \ = \  
    \begin{cases}
        \sin(z) \quad \ \text{for AdS branes,}\\
         \ \ \ z \quad \quad \ \text{for flat branes,}\\
        \sinh(z) \quad \text{for dS branes.}
    \end{cases}
\end{equation}
\begin{figure}[t]
    \centering
    \includegraphics[scale=0.57]{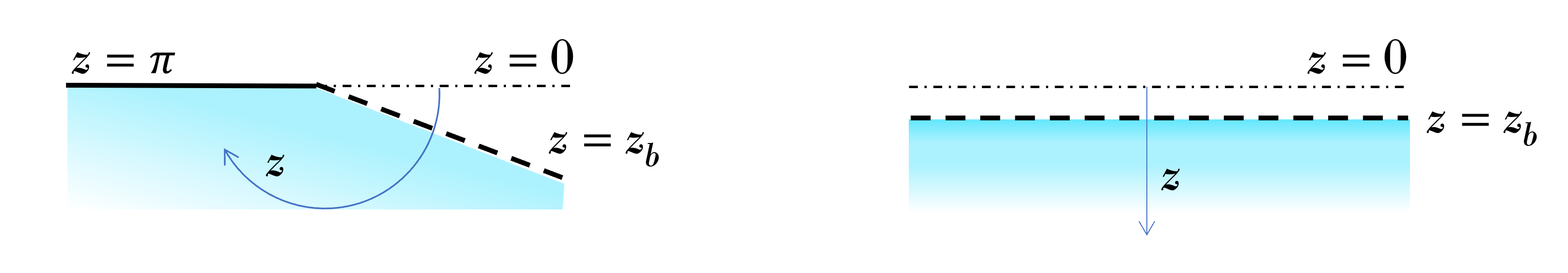}
    \caption{\small{Left: For AdS branes, the $z$ coordinate goes from $z = z_b$ to $z = \pi$. Right: For dS and flat branes, the $z$ coordinate goes from $z = z_b$ to the horizon at $z = \infty$. For flat branes, this is the usual bulk Poincaré horizon. For dS branes, the horizon is a Rindler horizon ---the brane must be accelerated in order not to fall into the AdS bulk.
    We will always work in the limit $z_b \to 0$.}}
    \label{fig:zcoords}
\end{figure}
Notice that the induced metric on constant $z$ slices is $g_{ij}(x,z) = \left(L/f(z)\right)^2 \hat{g}_{ij}(x)\,.$
Therefore, for (A)dS branes, the actual curvature radius of the induced brane geometry is 
\begin{equation}\label{lbrane}
    l^2 = \frac{L^2}{\left(f(z_b)\right)^2}\,.
\end{equation}

The unit normal vector to the brane is $\partial_n = - (f(z)/L) \partial_z$.
The minus sign comes from requiring that the normal vector be outward directed \cite{Chen:2020uac}. 
Then, the extrinsic curvature at constant $z$ slices is
\begin{align}
    K_{ij} & = \frac{f'(z)}{L} g_{ij}\,,
\end{align}
where prime denotes $z$-derivative.
The junction condition \eqref{IJC} at $z = z_b$ now reads
\begin{equation}\label{IJCTau}
    \tau = \tau_c f'(z_b) \,,
    \quad \quad \text{where} \quad \quad
    \tau_c := \frac{d-1}{8 \pi G_N L}\,.
\end{equation}
Since $\cos(z_b) < 1$, we must have $\tau < \tau_c$ for AdS branes, while $\tau > \tau_c$ corresponds to dS branes, with $\cosh(z_b) > 1$.
Flat branes have $\tau = \tau_c$ and their location is free.

It will be useful for later to rewrite the above metrics in a Fefferman-Graham fashion,
\begin{equation}\label{FGbulk}
    ds^2_{d+1} = \frac{L^2}{4\rho^2} d\rho^2 + \frac{L^2}{\rho} \tilde{g}_{ij}(\rho,x)dx^i dx^j\,,
\end{equation}
where
\begin{equation}\label{f(rho)Def}
    \tilde{g}_{ij}(\rho,x) = F(\rho) \hat{g}_{ij}(x)\,, \quad \quad \text{with} \quad \quad
    F(\rho) \ = \  
    \begin{cases}
        \left( \frac{1+\rho}{2} \right)^2 \quad \text{for AdS branes,}\\
        \quad \ 1 \quad \quad \ \ \text{for flat branes,}\\
        \left( \frac{1-\rho}{2} \right)^2 \quad \text{for dS branes.}
    \end{cases}
\end{equation}
Explicitly, the change of variables reads
\begin{equation}\label{fCoV-ztorho}
    \rho \ = \  
    \begin{cases}
        \ \tan^2 \left( \frac{z}{2} \right) \\
        \quad \ \ \ z^2\\
        \tanh^2 \left( \frac{z}{2} \right)
    \end{cases}
    \quad \text{and} \quad \quad
    f(z) \ = \  
    \begin{cases}
        \ \sin{z} = \frac{2\sqrt{\rho}}{1+\rho} \quad \quad \quad \text{for AdS branes,}\\
        \quad \ z \ = \sqrt{\rho} \quad \quad \quad \ \text{for flat branes\,,}\\
        \sinh z = \frac{2\sqrt{\rho}}{1-\rho} \quad \quad \quad \text{for dS branes.}
    \end{cases}
\end{equation}
The coordinate $\rho$ starts at $\rho_b \to 0$.
For the AdS case, the asymptotic boundary at $z = \pi$ is at $\rho = \infty$. For the flat and dS cases, the horizon at $z=\infty$ is at $\rho \to \infty$ and $\rho = 1$, respectively.

%%%%%%%%%%%%%%%%%%%%%%%%%%%%%%%%%%%%%%%%%%%%%%%%%%%%%%%%%%%%%%

%\bigskip

\subsection{Locally Localized Gravity - The Bulk Perspective}\label{sec:LLG}

Now, let us perturb the metric \eqref{SlicingMetric} with a linear axial transverse and traceless perturbation ($\delta G_{\mu z} = 0$, $\hat{g}^{ij} \delta \hat{g}_{ij} = 0$, $\hat{\nabla}^i \delta \hat{g}_{ij} = 0$),
\begin{equation}\label{PertSlicingMetric}
    d{s}^2_{d+1} = \frac{L^2}{\left(f(z)\right)^2} \left[ dz^2 + \left(\hat{g}_{ij}(x) + \delta \hat{g}_{ij}(x,z)\right) dx^i dx^j \right]\,,
\end{equation}
We are interested in these perturbations since they look like gravitons from the brane perspective. Moreover, the other potential modes are in fact pure gauge \cite{Karch:2001jb}.
From \eqref{EEsBulk}, we get \footnote{Details for these calculations can be found in Appendix B of \cite{LlorensGiralt:2024jaz}.}
%(see Appendix B of \cite{LlorensGiralt:2024jaz} for details for these calculations)
\begin{equation}\label{EEPert}
    \left[ \partial^2_z - (d-1) \frac{f'(z)}{f(z)} \ \partial_z + \left( \hat{\square} - 2 \sigma \right) \right]\delta \hat{g}_{ij}(x,z)  = 0,
\end{equation} 
where $\hat{\Box} = \hat{\nabla_i}\hat{\nabla}^i$, with $\hat{\nabla}$ being the Levi-Citiva connection of the unperturbed $\hat{g}_{ij}$ metric, and
\begin{equation}\label{sigma}
    \sigma \ := \  
    \begin{cases}
        -1 \quad \text{for AdS branes,}\\
         \ 0 \quad \ \text{for flat branes,}\\
        +1 \quad \text{for dS branes.}
    \end{cases}
\end{equation}
To linear order in perturbation, the extrinsic curvature on constant-$z$ hypersurfaces reads
\begin{align}
    \delta K_{ij} & = 
    \frac{L f'}{f^2} \delta \hat{g}_{ij} - \frac{f}{2L} \delta \hat{g}'_{ij}\,, &
    \delta K & = 0\,.
\end{align}
Then, using \eqref{IJCTau}, the junction condition \eqref{IJC} becomes a Neumann boundary condition 
\begin{equation}\label{BCPert}
    \left[ \partial_z \delta \hat{g}_{ij}(x,z) \right]_{z_b} = 0\,.
\end{equation}
Finally, assuming that the perturbation separates,
$\delta \hat{g}_{ij} (x,z) = H(z) h_{ij}(x)$,
and introducing the separation constant $E^2$, equation \eqref{EEPert} translates into
\begin{align}
     \left( \hat{\square} - 2 \sigma \right)h_{ij}(x) & = E^2 h_{ij}(x)\,, \label{EqBrane} \\
     \left[ \partial^2_z - (d-1) \frac{f'(z)}{f(z)} \ \partial_z \right] H(z) & = - E^2 H(z)\,, \label{EqRad}
\end{align}
while the boundary condition \eqref{BCPert} simply becomes
\begin{equation}\label{BCBrane}
    H'(z_b)=0\,.
\end{equation}

%%%%%%%%%%%%%%%%%%%%%%%%%%%%%

%\bigskip

\tocless\subsubsection{The Brane Equation}
Let us first study eq. \eqref{EqBrane}. This equation describes $\delta \hat{g}_{ij}(x,z)$ on hypersurfaces of constant $z$.
On the brane, we can actually rescale it in terms of the induced brane metric,
\begin{equation}\label{EqBrane2}
    \left( \square - \frac{2 \sigma}{l^2} \right)h_{ij}(x) = m^2 h_{ij}(x)\,,
\end{equation}
where $m^2 = E^2/l^2$ and $\Box = \nabla_i \nabla^i$, with $\nabla$ being the Levi-Citiva connection of the unperturbed induced metric on the brane $g_{ij}$. 
This equation describes a spin-2 massive mode in a maximally symmetric spacetime of radius $l$ (see \eg \cite{Bueno:2022lhf, Aguilar-Gutierrez:2023kfn}).
Therefore, from the perspective of the brane, indeed these $h_{ij}(x)$ perturbations look like massive spin-2 gravitons with mass $m^2$. 
We will be able to find the allowed masses $m^2$ by studying the radial equation \eqref{EqRad}.

%%%%%%%%%%%%%%%%%%%%%

%\bigskip

\tocless\subsubsection{The Radial Equation}\label{sec:EqRad}

Now we study the radial equation \eqref{EqRad} with boundary condition \eqref{BCBrane} on the brane.
This equation describes $\delta \hat{g}_{ij}(x,z)$ along the holographic direction $z$.
We will further impose Dirichlet boundary conditions on the asymptotic boundary on the other side of the spacetime
\begin{equation}
    H(z = \pi) = 0\,,
\end{equation}
for the AdS brane case, and regularity at the horizon at $z \to \infty$ for dS and flat branes.
We illustrate a couple of ways to solve this equation here (see Appendix C of \cite{LlorensGiralt:2024jaz} for more).

%%%%%%%%%%%

%\medskip

\begin{figure}[t]
    \centering
    \includegraphics[scale=0.51]{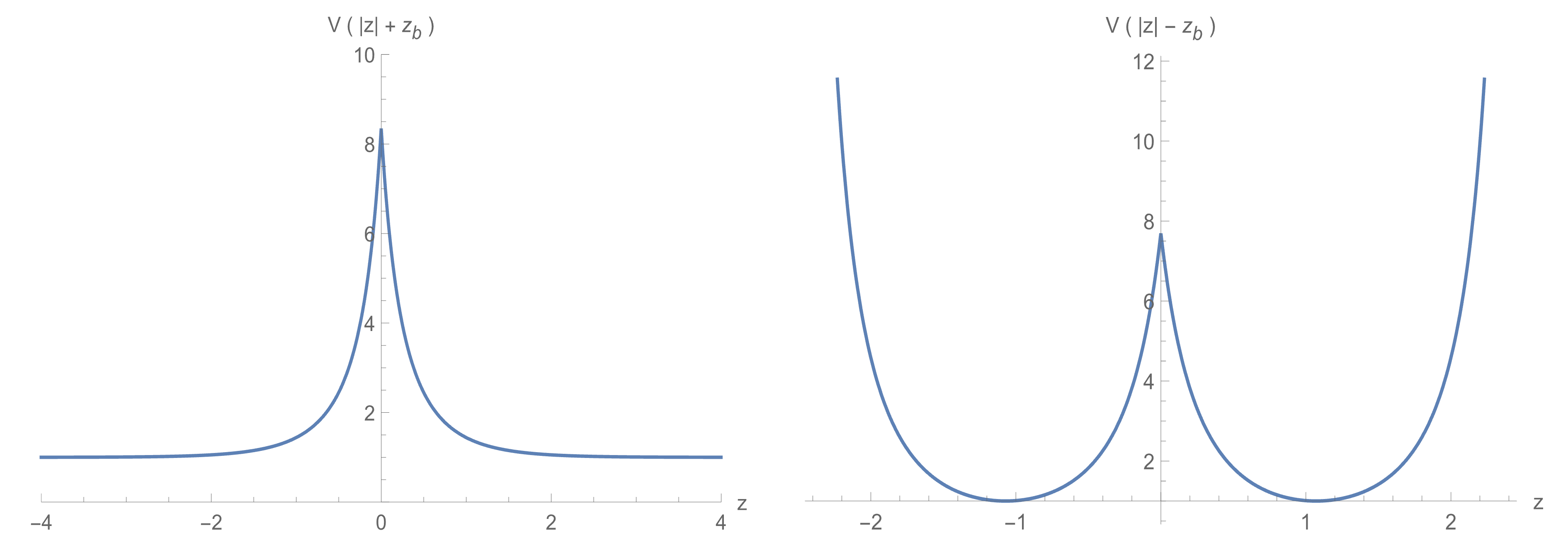}
    \caption{\small{Volcano $\mathbb{Z}_2$-symmetric $d=3$ potentials. Left: dS potential with $z_b = 1/2$. The flat case looks qualitatively the same, but decays to zero at infinity. Right: AdS potential with $z_b = \pi/8$.}}
    \label{VPotentials}
\end{figure}

\paragraph{The Volcano Potential.} 
Following \cite{Randall:1999vf,Karch:2000ct}, we redefine the radial function $H(z)$ as
\begin{equation}
   \tilde{H}(z) := \left( \frac{L}{f(z)} \right)^{\frac{d-1}{2}}H(z)\,,
\end{equation} 
to obtain a classical time-independent Schrödinger equation 
\begin{equation}\label{HRadSch}
    \left[- \partial_z^2 + V(z) \right] \tilde{H}(z) = E^2 \tilde{H}(z)\,,
\end{equation}
with potential
\begin{equation}
    V(z)  :=  \frac{d^2-1}{4}\frac{1}{\left(f(z)\right)^2}  + \sigma \frac{(d-1)^2}{4}\,.    
\end{equation}
From the shape of the potentials we see that the spectrum of eigenvalues will be continuous for flat and dS branes, since their potentials fall off at infinity (see Figure \ref{VPotentials}). On the other hand, the spectrum will be discrete for AdS branes, since the potential looks like a well. This is due to the Dirichlet boundary condition at $z = \pi$ boundary.

Moreover, after this redefinition, the brane boundary condition becomes
\begin{equation}
    \tilde{H}'(z_b) + \frac{d-1}{2}\frac{f'(z_b)}{f(z_b)}\tilde{H}(z_b) = 0\,,
\end{equation}
which acts as a delta function pointing downwards on the potential at $z = z_b$. This ensures the existence of a lowest-lying mode whose wave-function is localized on the brane. 

For flat and dS branes, the lowest-lying eigenvalue is exactly massless.
As seen from the falloff of the potentials at infinity, there is no mass gap between the lowest-lying eigenvalue and the continuum modes for flat branes, but there is a mass gap for dS branes, since the first excited state needs a minimum energy of $E^2_{\text{gap}} = (d-1)^2/4$.

For AdS branes, the lowest-lying mode is almost massless \cite{Karch:2000ct, Miemiec:2000eq, Schwartz:2000ip, Neuenfeld:2021wbl}, with mass
\begin{equation}
    E^2_0 \simeq \frac{d-2}{2^{d-1}} \frac{\Gamma(d)}{\left(\Gamma(d/2)\right)^2} z_b^{d-2}\,,
\end{equation}
while the excited modes have a mass of
\begin{equation}
    E^2_n = n(n+d-1) + \mathcal{O}(z_b^{d-2})\,.
\end{equation}

%%%%%%%%%

\paragraph{Full Solution.} Now that we have an idea of the problem at hand, let us solve it directly.

\begin{figure}[t]
    \centering
    \includegraphics[scale=0.58]{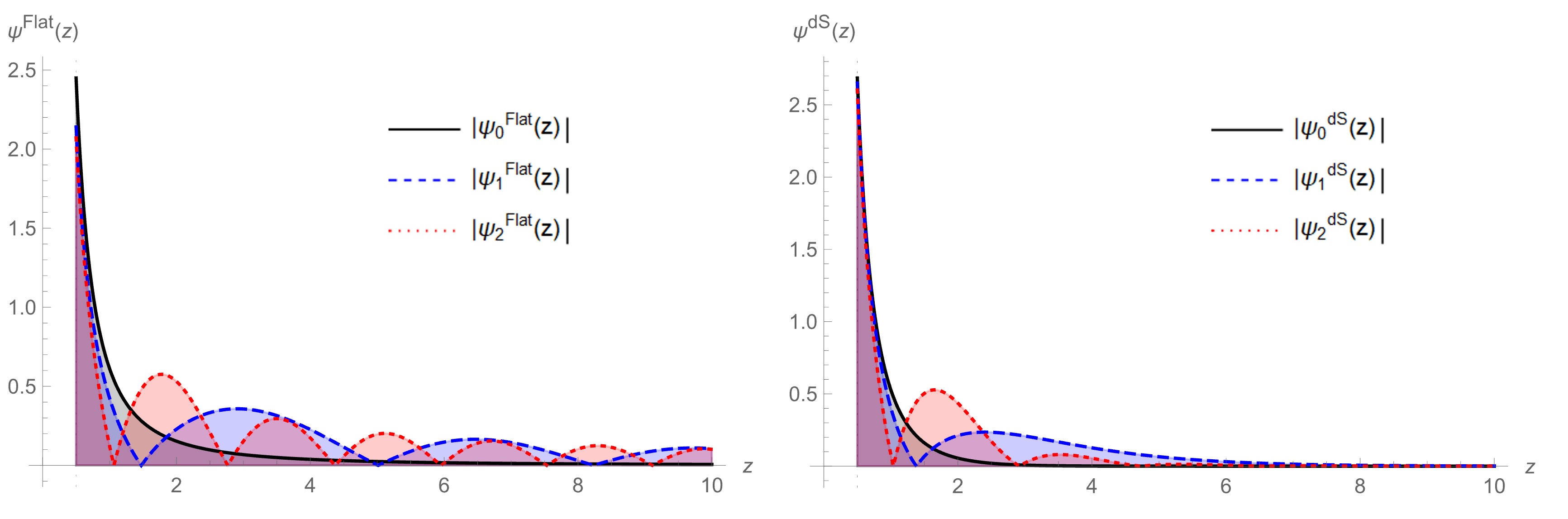}
    \caption{\small{Normalized radial profiles for a brane at $z_b = 1/2$, with $d = 3$, for $E^2 = 0$, $E^2 = 1$, and $E^2 = 4$. Left: Flat brane. Right: dS brane.}}
    \label{KKModes}
\end{figure}

First, we look for the massless mode on flat and dS branes. 
Solving the radial equation \eqref{EqRad} for $H(z)$ while imposing $E_0^2 = 0$ and the boundary condition \eqref{BCBrane} gives a constant value for $H(z)$, both for the dS and flat brane cases.
Recall however, from our definitions of the bulk perturbations in eqs. \eqref{PertSlicingMetric}, that once we take the warp factor into account, the actual radial profile of the bulk modes is $\psi(z) = \left(L/f(z)\right)^2 H(z)\,.$
Therefore, the radial profile of the zero mode goes as $\psi_0(z) \sim L^2/z^2$ for flat branes, and as $\psi_0(z) \sim L^2/\sinh{z}^2$ for dS branes, 
and so indeed the massless modes are localized on the brane (see Fig. \ref{KKModes}).

For flat branes, we can solve equation \eqref{EqRad} for the excited states of energy $E^2$ to find
\begin{equation}\label{HBessel}
    H(z) = c_1 z^{d/2} J_{d/2}(E z) + c_2 z^{d/2} Y_{d/2}(E z)\,,
\end{equation}
where $J, Y$ are respectively Bessel functions of the first and second kind, and $c_{1,2}$ are constants. 
For dS branes, we find that the excited states have the following radial profile,
\begin{align}\label{HHyper}
    H(z) = \ & c_1 \frac{\left(\sinh(z)\right)^d}{\left(\cosh(z)\right)^{1+\nu_+}} \prescript{}{2}{F_1}\left(\frac{1+\nu_+}{2},1+\frac{\nu_+}{2};1+\frac{d}{2};\tanh^2(z)\right) \\
    & + c_2 \frac{\left(\sinh(z)\right)^d}{\left(\cosh(z)\right)^{1+\nu_-}} \prescript{}{2}{F_1}\left(\frac{1+\nu_-}{2},1+\frac{\nu_-}{2};1-\frac{d}{2};\tanh^2(z)\right)\,.
\end{align}
where $\prescript{}{2}{F_1}$ is the hypergeometric function, again $c_{1,2}$ are arbitrary constants, and $\nu_\pm$ are
\begin{equation}
    \nu_\pm := \frac{\pm d - 1 + \sqrt{(d-1)^2-4E^2}}{2}\,.
\end{equation}
In both cases, imposing the boundary condition \eqref{BCBrane} on the brane fixes the ratio of the constants $c_1/c_2$ as a function of the position of the brane $z_b$.

The solution to equation \eqref{EqRad} for AdS branes is
\begin{equation}\label{HLegendrec1c2}
H(z) = c_1 (\sin z)^{\frac{d}{2}}P^{d/2}_\nu(\cos z) + c_2 (\sin z)^{\frac{d}{2}}Q^{d/2}_\nu(\cos z)\,,
\end{equation}
where $P^{d/2}_\nu$ and $Q^{d/2}_\nu$ are associated Legendre polynomials, $\nu$ is defined as
\begin{equation}\label{Defnu}
    \nu := \frac{-1 + \sqrt{(d-1)^2+4E^2}}{2}\,,
\end{equation}
and $c_1$ and $c_2$ are complex arbitrary constants \cite{Neuenfeld:2021wbl}.
Imposing $H(z = \pi) = 0$, we find
\begin{equation}\label{c2c1}
    \frac{c_2}{c_1} = \frac{2}{\pi} \cot \left( \pi \nu + \frac{\pi}{2} \right)\,.
\end{equation}
Further imposing the boundary condition \eqref{BCBrane} on the brane discretizes the spectrum. This can be done numerically, or analytically in the limit $z_b \to 0$. In any case, it is easier to do so in Fefferman-Graham coordinates \eqref{fCoV-ztorho}. The boundary condition on the brane then reads
\begin{equation}\label{BCBraneRho}
    \left[ \partial_\rho H(\rho) \right]_{\rho_b} = 0\,.
\end{equation}
We will study the eigenvalues $E^2$ as a function of the brane position $\rho_b$, and not its tension $\tau < \tau_c$. Recall that both quantities are directly related by eq. \eqref{IJCTau}, and that tuning the tension close to its critical value, $\tau \to \tau_c$, brings the brane closer to the boundary, $\rho_b \to 0$.

\begin{figure}[t]
    \centering
    \includegraphics[scale=0.46]{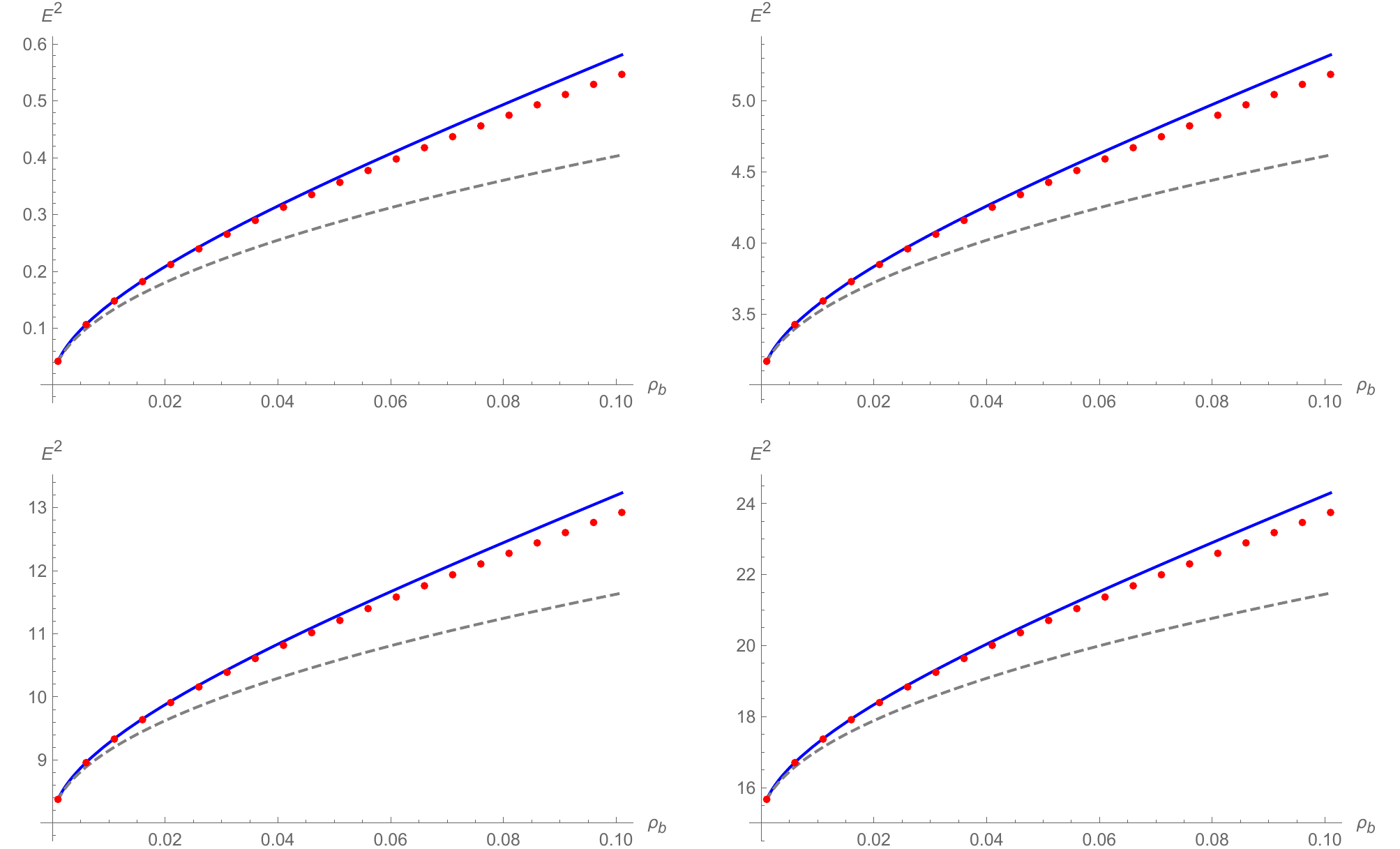}
    \caption{\small{Lowest-lying eigenvalues $E^2_n$ for $d=3$ as a function of the brane position $\rho_b$. Red dots are numerical results, blue lines correspond to the improved analytical approximation in \eqref{E2nd3}, while gray dashed lines correspond to the analytical approximation in \eqref{E2nd}. Top-left: almost-massless mode, with $n = 0$. Top-right: $n = 1$. Bottom-left: $n = 2$. Bottom-right: $n = 3$.
    }}
    \label{Eigen}
\end{figure}

Analytically, in the limit $\rho_b \to 0$, we find
\begin{equation}\label{E2nd}
    E^2_{(n,d)} \simeq n(n+d-1) + \frac{1}{2}(d-2)(2n+d-1) \frac{\Gamma(n+d-1)}{\Gamma(n+1)(\Gamma(d/2))^2} \rho_b^{d/2-1}\,,
\end{equation}
where $n = 0,1,2,\ldots$ . We can see that as the brane is sent to the boundary, that is, as $\rho_b \to 0$, we recover the usual eigenvalues for the graviton modes of global AdS$_{d+1}$,
\begin{equation}\label{E2AdS}
    E^2_{(n,d)} (\rho_b = 0) = n(n+d-1)\,.
\end{equation}
This fact is due the Neumann-like boundary condition \eqref{BCBrane} for the radial equation becoming a Dirichlet boundary condition at infinity: the brane is allowed to fluctuate, but as we send it to the boundary by increasing its tension, it becomes stiffer. In the limit where the brane reaches the boundary, it is as if it were infinitely stiff, so we recover the usual Dirichlet boundary conditions used in standard AdS/CFT.

To find the analytical expansion \eqref{E2nd} above, we expanded $H'(\rho)$ in terms of $u = \sqrt{\rho}$ around $u = 0$ to ($d-2$) order,and again in terms of $E^2_{(n,d)}$ around $n(n+d-1)$ to linear order. Then, we solved for $E^2_{(n,d)}$ after imposing the boundary condition \eqref{BCBraneRho}.
With this method, we have actually been able to find a better approximation for the $d=3$ eigenvalues, 
\begin{equation}\label{E2nd3}
    E^2_{(n, d=3)} \simeq \frac{n(n+2)\pi + (n^2+2n+4)\sqrt{\rho_b}}{\pi - 3\sqrt{\rho_b}}\,.
\end{equation}
We have also constructed similar expressions for $d=5$ and $d=7$ case by case in $n$ (see Appendix C of \cite{LlorensGiralt:2024jaz}), but we could not find an improved formula for all $(n,d)$.

Now, if $\rho_b \to 0$, we have $2\sqrt{\rho_b} \simeq z_b $. Therefore, the graviton masses in \eqref{EqBrane2} are
\begin{equation}\label{GravMass}
    m^2_{(n,d)} \simeq n(n+d-1)\frac{z_b^{2}}{L^2} + \frac{(d-2)(2n+d-1)}{2^{d-1}} \frac{\Gamma(n+d-1)}{\Gamma(n+1)(\Gamma(d/2))^2} \frac{z_b^{d}}{L^2}\,.
\end{equation}
In particular, this agrees with the results found in \cite{Neuenfeld:2021wbl} for the mass of the lowest-lying mode.

\begin{figure}[ht]
    \centering
    \includegraphics[scale=0.65]{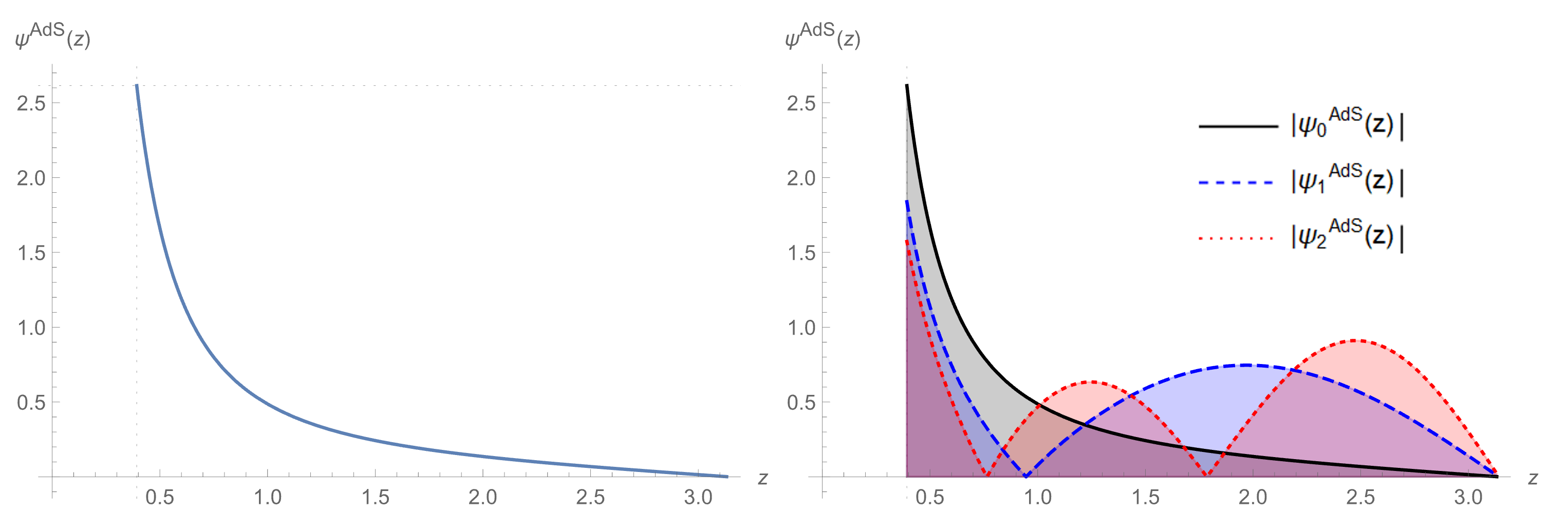}
    \caption{\small{Normalized radial profiles for various modes on a $d=3$ AdS brane at $z_b = \pi/8$. Left: Normalized radial profile for the $n=0$ mode. Right: Comparison with the $n = 1$ and $n = 2$ modes.}}
    \label{ZeroModesAdS}
\end{figure}

Again, the lowest-lying mode is localized on the brane, since its radial behaviour goes as $\psi_0(z) \sim 1/\sin^2(z)$ for $|z - z_b| \ll 1$ when the brane is close to the boundary (see Fig. \ref{ZeroModesAdS}).

%%%%%%%%%%%%%%%%%%%%%%%%%%

\tocless\subsubsection{Gravity on the Brane}

For flat $d = 4$ branes, Randall and Sundrum already showed that the massless mode gave the expected Newtonian potential on the brane \cite{Randall:1999vf}.
Taking the continuum of Kaluza-Klein modes into account, Garriga and Tanaka \cite{Garriga:1999yh} proved that four-dimensional Randall-Sundrum brane-worlds display a potential at a distance $r \gg L$ around a point-like mass $M$ of the form
\begin{equation}\label{NewtVBrane}
    V(r) = \frac{G_{N,\text{eff}} M}{r} \left( 1 + \frac{2 L^2}{3 r^2} \right)\,,
%\end{equation}
%for $r \gg L$, and 
\quad \quad
\text{where} 
\quad \quad
%\begin{equation}
    G_{N,\text{eff}} := \frac{2}{L}G_{N}\,.
\end{equation}
Therefore, at long distances, gravity on the brane becomes four-dimensional. However, gravity remains higher-dimensional at distances shorter than the bulk curvature radius $L$.

Later on, by rearranging the bulk Einstein equations, Shiromizu, Maeda, and Sasaki \cite{Shiromizu:1999wj} proved that one can covariantly obtain the full Einstein equations on the brane.
Their result, however, might be misleading if not interpreted with care.
There appears an extra term on the matter side of the four-dimensional Einstein equations, the projection of the bulk Weyl tensor on the brane, which captures the effects of five-dimensional gravity.
It is the non-linear generalisation of the linear Kaluza-Klein modes.
From the perspective of a brane observer, these effects cannot be determined purely from brane data, since they come from the full five-dimensional bulk.
Using the holographic duality, we will reinterpret these terms as the CFT radiation dual to the AdS bulk.

The results above can be generalized to any number of spacetime dimensions, now with
\begin{equation}
    G_{N,\text{eff}} := \frac{d-2}{L}G_{N}\,,
\end{equation}
and to (A)dS branes, provided that they sit close enough to the asymptotic boundary so that their geometry is almost flat.
This generalization relies on the behaviour of the linear graviton modes being dominated by the warp factor $1/\left(f(z)\right)^2$ for $z \ll 1$, and that $f(z)$ behaves as $f(z) \sim z$ for $z \ll 1$ regardless of the brane geometry. 

%%%%%%%%%%%%%%%%%%%%%%%%%%%%%%%%%%%%%%%%%%%%%%%%%%%%%%%
%
%
%
%%%%%%%%%%%%%%%%%%%%%%%%%%%%%%%%%%%%%%%%%%%%%%%%%%%%%%%

\bigskip

\subsection{Induced Gravity on the Brane - The Brane Perspective}\label{sec:BranePOV}

We will now make use of the holographic duality to get an understanding of the brane physics solely in terms of brane quantities, encoding the bulk as the stress-energy tensor of a strongly coupled CFT on the brane \cite{deHaro:2000wj, Neuenfeld:2021wbl}.
As in standard AdS/CFT, this quantity is proportional to a particular coefficient of the bulk metric written in a Fefferman-Graham expansion \cite{Skenderis:2002wp},
so one can compute it directly if the bulk metric is known.
Alternatively, given a metric on the brane, we can define the CFT stress-energy tensor as the right-hand-side of the brane Einstein equations.
Then, this must be made consistent by solving the bulk boundary problem to find a bulk metric fulfilling the bulk Einstein equations with the required boundary conditions on the brane \cite{deHaro:2000wj}. 
Although feasible, it is usually more practical to
work with known bulk metrics, as in the C-metric papers describing quantum black holes on branes (see \eg \cite{Emparan:1999wa, Emparan:1999fd, Emparan:2000rs, Emparan:2022ijy, Emparan:2023dxm, Feng:2024uia, Climent:2024nuj, Climent:2024wol, Panella:2023lsi, Panella:2024sor, Cartwright:2025fay}).

\tocless\subsubsection{The Brane Effective Action}
\label{subsec:IeffBrane}

One can integrate the bulk following a ``finite'' holographic renormalization prescription \cite{Neuenfeld:2021wbl, deHaro:2000wj, Skenderis:2002wp, Bueno:2022log} to obtain an effective description of the brane dynamics written purely in brane variables. The result is the following effective action on the brane,
\begin{equation}\label{Ieff}
    I_{\text{eff}} = I_\text{bgrav} + I^{UV}_\text{CFT}\,.
\end{equation}
The term $I^{UV}_\text{CFT}$ is dual to the AdS$_{d+1}$ bulk and describes a holographic CFT, which has a UV cut-off because the bulk ends at the brane, at some finite distance from the boundary.
The term $I_\text{bgrav}$ is an effective higher-derivative theory of gravity on the brane,
\begin{equation}\label{Ibgrav}
     I_\text{bgrav} = \frac{1}{16 \pi G_{N,\text{eff}} } \int \df^d x \sqrt{-g} \left[R-2\Lambda_{\text{eff}} + \frac{L^2}{(d-4)(d-2)}( R^{ab}R_{ab} - \frac{d}{4(d-1)} R^2) + \cdots \right]\,,
\end{equation}
where all curvature tensors are built from the induced metric on the brane, and
\begin{align}\label{GdLd}
    G_{N,\text{eff}} & := \frac{d-2}{L}G_{N}\,, & \Lambda_{\text{eff}} & := - \frac{(d-1)(d-2)}{L^2} \left( 1 - \frac{\tau}{\tau_c} \right)\,.    
\end{align}
This term $I_\text{bgrav}$ arises from solving the bulk Einstein equations in the near-boundary region excluded by the brane.
How? 
First, we need to review standard holographic renormalization.

In conventional AdS/CFT, the bulk partition function diverges, since bulk distances diverge near the boundary. 
These are long-distance (IR) divergences in the bulk, but correspond to UV divergences in the CFT.
To remove them and obtain a useful, finite, bulk partition function which we can equate to the CFT partition function, one must add counterterms,
\begin{equation}\label{Ifinbulk}
    I_\text{bulk}^\text{fin} = I_\text{bulk} + I_{\text{ct}}\,.
\end{equation}
These counterterms can be written as local curvature tensors of the boundary metric \cite{Skenderis:2002wp, Kraus:1999di, deHaro:2000vlm, Emparan:1999pm}.

Now, since our spacetime ends on the brane, at some finite distance $z_b$ from the boundary, the on-shell action no longer diverges.\footnote{
In the case of AdS branes, the bulk reaches the asymptotic boundary on the side of the spacetime far from the brane, at $z = \pi$. There the on-shell action diverges, and we must add the usual counterterms.}
If the brane is sufficiently close to the boundary, however, the dependence of $I_\text{bulk}$ with $z_b$ will have the same structure as the counterterms $I_\text{ct}$, since they just reflect the AdS asymptotics.
We collect these finite ``counterterms'' under $I^{z_b}_{\text{ct}}$.
The terms at order $n < d$ in metric derivatives would diverge as the brane reaches the boundary, while the terms at order $n > d$ would vanish.
The term at order $n = d$ would give rise to the trace anomaly in standard AdS/CFT \cite{Henningson:1998gx} for spacetimes with an even number of boundary dimensions.
However, since the brane sits at a finite distance from the boundary, all terms are finite, so $I^{z_b}_{\text{ct}}$ contains an infinite tower of higher-derivative terms.

It is now useful to add and subtract these finite ``counterterms'' $I^{z_b}_\text{ct}$ \cite{Neuenfeld:2021wbl}, to write \eqref{ActionBulk} as
\begin{equation}
    I_\text{bulk} + I_\text{brane} = \left( I_\text{bulk} + I^{z_b}_{\text{ct}}\right) + \left(I_\text{brane} - I^{z_b}_{\text{ct}}\right) \,.
\end{equation}
We can now identify the first term as $I^{UV}_\text{CFT}$,
\begin{equation}\label{IUVCFT}
    I^{UV}_\text{CFT} = I_\text{bulk} + I^{z_b}_{\text{ct}}\,,
\end{equation}
a CFT with a cut-off given by the distance from the brane to the boundary. Indeed, if we were to send $z_b \to 0$, this first term would become $I_\text{bulk}^\text{fin}$, as in eq. \eqref{Ifinbulk}, which we would translate into the standard CFT partition function using the holographic dictionary.

The remaining term is then the effective gravitational action for the brane dynamics,
\begin{equation}
    I_\text{bgrav} = I_\text{brane} - I^{z_b}_{\text{ct}}\,,
\end{equation}
whose explicit expression is given in \eqref{Ibgrav} above. 
Notice that $I^{z_b}_{\text{ct}}$ contains the tower of higher-derivative operators seen in eq. \eqref{Ibgrav}, with the structure of the standard counterterms, while $I_\text{brane}$ is simply the brane tension \eqref{IBrane}, which tunes the cosmological constant $\Lambda_\text{eff}$, as shown in eq. \eqref{GdLd}.
Moreover, $I_{\text{bgrav}}$ is large when $z_b$ is small, since it is mostly $-I_\text{ct}^{z_b}$, whose first terms would diverge as $z_b \to 0$. 
This shows the strong localization of gravity on the brane.

If $\tau < \tau_c$, the cosmological constant $\Lambda_\text{eff}$ on the effective action is negative, so we will have AdS asymptotics on the brane, while if $\tau > \tau_c$, then the effective cosmological constant is positive, and so the brane will have dS asymptotics. 
As excepted, $\Lambda_\text{eff}$ becomes zero when $\tau = \tau_c$, in accordance with our results from the previous section.
We can rewrite it as 
\begin{equation}
    \Lambda_{\text{eff}} = \sigma \frac{(d-1)(d-2)}{2 \ell^2}\,,
\end{equation}
where $\sigma$ again denotes the sign of the brane, and we have defined
\begin{equation}\label{ellbrane}
    \ell^2 := \frac{L^2}{2} \left| 1 - \frac{\tau}{\tau_{c}} \right|^{-1}\,.
\end{equation}
For (A)dS branes, in the limit $\tau \to \tau_c$, if we define $\varepsilon := \left| 1 - \tau/\tau_c \right|/2 \ll 1$,
we have
\begin{equation}
    \frac{L^2}{\ell^2} = 2 \left|1 - \frac{\tau}{\tau_c} \right| = 4\varepsilon\,.
\end{equation}
This makes it clearer that indeed the action \eqref{Ieff} is an effective action, with each higher-derivative term parametrically smaller than the previous one \cite{Emparan:2020znc, Emparan:2022ijy}.

Although $\ell$ looks very similar to the brane curvature radius $l$, they only match to linear order in $\varepsilon$. Indeed, for (A)dS branes, we can write the radius $l$ defined in eq. \eqref{lbrane} as
\begin{equation}
    l^2 
    = L^2\left|1 - \left(\frac{\tau}{\tau_c}\right)^2\right|^{-1}\,,
\end{equation}
which in the limit $\tau \to \tau_c$ exactly reads $L^2/l^2 = 4\varepsilon + 4\varepsilon^2$.

Making use of equations \eqref{IJCTau} and \eqref{fCoV-ztorho}, it is easy to see that the position of the brane is also controlled by this same parameter $\varepsilon$. In Poincaré-like coordinates \eqref{SlicingMetric}, it is
\begin{equation}
    z_b = (f')^{-1}\left(\tau/\tau_c\right) \simeq 2\sqrt{\varepsilon}\ + \mathcal{O}(\varepsilon)\,,
\end{equation}
while in Fefferman-Graham coordinates \eqref{FGbulk}, we have
\begin{equation}
    \rho_b = \left| \frac{\tau_c-\tau}{\tau_c+\tau} \right| \simeq \varepsilon\ + \mathcal{O}(\varepsilon^2)\,.
\end{equation}

%%%%%%%%%%%%%%

\tocless\subsubsection{Graviton Mass from the Brane Perspective}
\label{subsec:GravMassBranePOV}

We will now reinterpret, from a brane-world holographic perspective, how the graviton on AdS branes acquires its mass.
One might be tempted to say that the mass of the graviton ---and similarly, the mass of the Kaluza-Klein modes--- comes from the higher-derivative operators in \eqref{Ieff}.
After all, higher-derivative terms generally induce massive modes in the spectrum when linearizing around a maximally symmetric spacetime (see \eg \cite{Bueno:2022lhf, Aguilar-Gutierrez:2023kfn}).
However, the contributions to the mass coming from the higher-derivative terms enter at a different order in $L$ and can only account for corrections.
Instead, the main contribution to the mass of the lowest-lying graviton on AdS branes comes from the interaction between gravity and CFT on the brane.
To see this, we will linearize the equations of motion of the effective brane action \eqref{Ieff}, and then relate each side of the equations to bulk quantities while making use of our results from the previous Section \ref{sec:LLG}.
So we will not find a \textit{new} way to compute the mass of the graviton, but a way to reinterpret the bulk results from the brane perspective.
We will closely follow \cite{Neuenfeld:2021wbl} in this subsection, where one can find a more detailed explanation of this topic.

Varying the brane effective action \eqref{Ieff}, we obtain
\begin{equation}\label{BraneEEs}
    R_{ab} - \frac{1}{2} R g_{ab} + \Lambda_\text{eff} \ g_{ab} + \cdots = 8 \pi G_{N,\text{eff}} T^{\text{CFT}}_{ab} \,,
\end{equation}
where $T^{\text{CFT}}_{ab}$ is the stress-energy tensor obtained from varying $I^{UV}_\text{CFT}$ with respect to the induced brane metric $g_{ab}$, and the ellipses denote the EOMs of the higher-derivative terms in $I_\text{bgrav}$.

Since $I^{UV}_\text{CFT}$ is given by \eqref{IUVCFT}, one can adapt the standard AdS/CFT results to argue that, to leading order, the CFT stress-energy tensor on the brane is \cite{Neuenfeld:2021wbl}
\begin{equation}\label{TijCFT}
    \langle T^{\text{CFT}}_{ij} \rangle = \varepsilon^{d/2-1} \left( \frac{d L}{16 \pi G_N} \tilde{g}^{(d)}_{ij} + X^{(d)}_{ij} \left[ \tilde{g}^{(0)} \right] \right)\,,
\end{equation}
where $\tilde{g}^{(0)}_{ij}$ and $\tilde{g}^{(d)}_{ij}$ are the terms that appear, respectively, at order $\rho^0$ and $\rho^{d/2}$ in the Fefferman-Graham (FG) expansion of the bulk metric on the brane,\footnote{The term $X^{(d)}_{ij}$ in eq. \eqref{TijCFT} and the coefficient $\tilde{h}^{(d)}_{ij}$ in \eqref{FGg} only appear for even $d$ \cite{Fefferman:2007rka}.
Through the bulk Einstein equations, they are fixed in terms of $\tilde{g}^{(0)}_{ij}$, and give raise to the trace anomaly of $\langle T^{\text{CFT}}_{ij} \rangle$ \cite{Henningson:1998gx}.}
\begin{equation}\label{FGg}
    G_{ij}(\rho,x) = \frac{L^2}{\rho} \left( \tilde{g}^{(0)}_{ij} + \tilde{g}^{(2)}_{ij} \rho + \cdots + \tilde{g}^{(d)}_{ij} \rho^{d/2} + \tilde{h}^{(d)}_{ij}  \rho^{d/2} \log (\rho) + \mathcal{O}(\rho^{d/2+1}) \right) \,.
\end{equation}

Now, in the limit $\tau \to \tau_c$, the brane Einstein equations \eqref{BraneEEs} up to order $n$ in curvature are solved, up to order $\varepsilon^{n+1}$, by a vacuum AdS$_d$ metric with curvature radius $l$, even though we saw that $\ell$ and $l$ only coincide to linear order in $\varepsilon$. This is because the higher-curvature terms act as extra contributions to the cosmological constant, bringing $\ell$ and $l$ together.
Then, perturbing the brane metric $g_{ij} \to g_{ij} + \delta g_{ij}$, we obtain
\begin{equation}\label{LinEqBrane}
    \left( \Box + \frac{2}{l^2} + \cdots \right) \delta g_{ij} = -16\pi G_{N,\text{eff}} \ \delta T^\text{CFT}_{ij}.
\end{equation}
But this CFT stress-energy tensor is not an arbitrary stress-energy tensor ---we just saw that it is proportional to the $\tilde{g}^{(d)}_{ij}$ coefficient of the bulk metric expressed in FG coordinates. Similarly, the induced metric on the brane is proportional to the $\tilde{g}^{(0)}_{ij}$ term of the bulk metric in the FG expansion. 
Therefore, we can relate $\delta g_{ij}$ and $\delta T^\text{CFT}_{ij}$, making use of the expansion \eqref{FGg} and the following expressions for the brane metric perturbations
\begin{equation}
    \delta g_{ij}(x) = \delta G_{ij} (\rho_b,x) = \frac{L^2}{\rho_b} H(\rho_b) h_{ij}(x)\,,
\end{equation}
to write \cite{Neuenfeld:2021wbl},
\begin{equation}
    \left.\begin{cases}
        \ \tilde{g}^{(d)}_{ij} (x) = B_0 h_{ij} (x) \\
        \ \delta g_{ij}(x) = \frac{L^2}{\rho_b} A_0 h_{ij} (x) + \mathcal{O}(\rho_b)
    \end{cases} \right\}
    \implies \tilde{g}^{(d)}_{ij} (x)
    \simeq \frac{\rho_b}{L^2} \frac{B_0}{A_0} \delta g_{ij} (x)\,,
\end{equation}
where $B_0$ is the coefficient of the term $\propto \rho^{d/2}$ and $A_0$ is the coefficient of the term $\propto \rho^{0}$ of $H(\rho)$ when expanded close to $\rho_b \to 0$.
Finally, substituting all these results into eq. \eqref{LinEqBrane}, and relating again the bulk and brane Newton's constant through eq. \eqref{GdLd}, we obtain \cite{Neuenfeld:2021wbl}  
\begin{equation}\label{LinEqBrane2}
    \left( \Box + \frac{2}{l^2} + \cdots \right) \delta g_{ij} = -d(d-2) \frac{B_0}{A_0}\frac{\varepsilon^{d/2}}{L^2} \delta g_{ij} + \cdots\,.
\end{equation}
Neglecting the higher-derivative terms, we see that this equation above is an equation for a massive graviton, and is, in fact, the rescaled version of the brane equation \eqref{EqBrane2}, if we identify
\begin{equation}\label{GravMassBrane}
    m^2_0 = -d(d-2)\frac{B_0}{A_0}\frac{\varepsilon^{d/2}}{L^2}\,.
\end{equation}
Therefore, we will be able to find the mass of the graviton on AdS branes if we can expand the bulk metric à la Fefferman-Graham and find the coefficients $A_0$ and $B_0$.

Indeed, in the previous section, we found that the radial profile of bulk perturbations $\psi(\rho) \sim L^2 H(\rho)/\rho$ could be written as a superposition of associated Legendre polynomials.
Near the brane at $\varepsilon \to 0$, these Legendre polynomials can be expanded into \cite{NIST:DLMF}, 
\begin{equation}\label{HlimRhob}
    H_n(\rho) \ \simeq
    \begin{cases}
        -2^{d/2} \Bigg[(-1)^{\frac{d-1}{2}}\frac{\pi \cos(\nu\pi)}{2\Gamma(1-d/2)} +\frac{\pi \sin(\nu\pi) \Gamma(\nu+d/2+1)}{2\Gamma(d/2+1)\Gamma(\nu-d/2+1)} \rho^{d/2} \Bigg] + \cdots \ \ \quad \quad \quad \quad \quad \text{for odd $d$,}\\
        - \frac{\sin(\pi \nu)}{\pi} 2^{d/2} \left[ \Gamma(d/2) + (-1)^{d/2} \cos(\nu\pi) \frac{\Gamma(d/2-\nu)\Gamma(d/2+\nu+1)}{\Gamma(d/2+1)} \rho^{d/2} \right] + \cdots  \quad \text{for even $d$,}
    \end{cases}
\end{equation}
where $\nu$ was defined in eq. \eqref{Defnu}. 

We can now read out the quotient
\begin{equation}
    \frac{B_n}{A_n} =
    \begin{cases}
         (-1)^{\frac{d-1}{2}} \sin(\nu\pi) \frac{\Gamma(d/2-\nu)\Gamma(d/2+\nu+1)}{\Gamma(d/2)\Gamma(d/2+1)}\,,  \quad \text{for odd $d$,}\\
         (-1)^{d/2} \cos(\nu\pi) \frac{\Gamma(d/2-\nu)\Gamma(d/2+\nu+1)}{\Gamma(d/2)\Gamma(d/2+1)}\,,\quad \text{for even $d$.}
    \end{cases}
\end{equation}
Assuming as ansatz that the lowest-lying eigenvalue behaves as $E_0^2 \to 0$ as $\varepsilon \to 0$, we see $\nu \to \frac{d}{2} - 1$,
so regardless of the number of brane dimensions $d$,
\begin{equation}\label{BoverA}
    \frac{B_0}{A_0} \simeq -\frac{2}{d}\frac{\Gamma(d)}{(\Gamma(\frac{d}{2}))^2}\,.
\end{equation}
Plugging this into equation \eqref{GravMassBrane} for $m_0^2$, we finally obtain
\begin{equation}
    m^2_0 = \frac{2(d-2)\Gamma(d)}{(\Gamma(\frac{d}{2}))^2}\frac{\varepsilon^{d/2}}{L^2}\,,
\end{equation}
which coincides with our previous results, as seen in equation \eqref{GravMass}, with $2 \sqrt{z_b} \simeq \varepsilon$. 

Finally, a word on why the graviton on flat and dS branes is massless. Recall that the zero mode for flat and dS branes had constant $H(z)$, and so it had radial profile,
\begin{equation}
    \psi(\rho) = \frac{1}{\rho} H(\rho) \propto \frac{1}{\rho}\,.
\end{equation}
This means that the zero mode contains only a non-normalizable piece. Therefore, it does not contribute to the brane CFT stress-energy tensor, and, following a similar argument as the one above for AdS branes, we would see that it is massless.

%%%%%%%%%%%%%%%%%%%%%%%%%%%%%%%%%%%%%%%%%%%%%%%%%%%%%%%%%%%%%%%%%%%%%%%%%%%%%%%%%%%%%%%%%%%%%
%
%
%
%
%
%%%%%%%%%%%%%%%%%%%%%%%%%%%%%%%%%%%%%%%%%%%%%%%%%%%%%%%%%%%%%%%%%%%%%%%%%%%%%%%%%%%%%%%%%%%%%

\bigskip

\section{Brane-Worlds with DGP Terms}
\label{chp:BWsWithDGP}

Soon after the RS paper, Dvali, Gabadadze and Porrati \cite{Dvali:2000hr} presented an alternative way to obtain four-dimensional gravity on a brane in a five-dimensional spacetime, considering a flat brane with an explicit Einstein-Hilbert term sitting on a Minkowski bulk,
\begin{equation}
    I_{\text{DGP}} = \frac{1}{16 \pi G_N} \left( \int_\text{Bulk} \df^5 x \sqrt{-G} \ R[G] + \lambda \int_\text{brane} \df^4x \sqrt{-g} \ R \right)\,,
\end{equation}
where $\lambda$ is some length scale.
Unlike the KRS models, their work recovered four-dimensional gravity at short scales but not at distances larger than $\lambda$ \cite{Luty:2003vm} ---gravity leaks off the brane into the bulk at large scales. 
Generalizing DGP models to allow for FLRW geometries on the brane \cite{Deffayet:2000uy}, this large-scale weakening of gravity induces an accelerated expansion of the brane geometry \cite{Maartens:2010ar}.
Unfortunately, it was later discovered that DGP models with expanding branes are theoretically unstable, since the scalar sector of gravitational perturbations contains an infrared ghost \cite{Charmousis:2006pn, Koyama:2007za}, so DGP models were ruled out as a model for our Universe \cite{Maartens:2010ar}.

However, we are interested in holography, not cosmology, so we will consider DGP branes sitting on an AdS bulk.
From an EFT point of view, after the tension, the DGP term is the next natural term in an effective expansion of the brane action. Thus, this is a straightforward generalisation of the KRS models studied in the previous section.
We will allow for all three possible brane maximally symmetric geometries, and study the brane location as a function of the DGP coupling.
Then, we will explore how the localization of gravity on the brane changes due to the DGP term.
In particular, we will look for the presence of pathologies in the theory, which will put a bound on the allowed values for the DGP coupling.
We will see that positive values for the DGP coupling are always allowed, as well as a small enough negative coupling.
However, we cannot have large negative DGP couplings, or the whole construction breaks down: either the brane moves to a location where gravity no longer localizes, or we get undesired graviton modes on the brane.
Finally, we will again reinterpret these results from the perspective of the dual brane picture.
Throughout this whole section, our emphasis will be on the AdS brane case, since it is the one that is most relevant for research in holography.

%%%%%%%%%%%%%%%%%%%%%%%%%%%%%%%%%%%%%%%%

\subsection{Set-up and Brane Position}
\label{sec:SetUpDGP}

Our set-up is again an AdS$_{d+1}$ bulk with radius $L$ ending on a codimension-one brane, as described in \eqref{IBulk}, but we now add an explicit Einstein-Hilbert term on the brane,
\begin{equation}
    I_{\text{brane}} = \int_{\partial\mathcal{M}_b} \df^d x \sqrt{-g} \left( -\tau + A \frac{L}{8 \pi G_N (d-2)} R \right)\,,
    \label{IbrDGP}
\end{equation}
where $R$ is the Ricci scalar built from the brane induced metric $g_{ab}$, and we have chosen to normalize the DGP coupling constant $A$ in this way for simplicity in future calculations.

Since the DGP term is only present on the brane, the bulk Einstein equations \eqref{EEsBulk} remain unchanged. However, the Israel junction condition on the brane now reads
\begin{equation}\label{IJCDGP}
    K_{ab} - K g_{ab} = - 8 \pi G_N \tau g_{ab} + A\frac{L}{d-2} \left( R g_{ab} - 2R_{ab} \right)\,,
\end{equation}
where all tensors are built from the induced metric $g_{ab}$.

%%%%%%%%%%%%%%%%%%%%%%%%%%%%%%%%%%%%%%%%

Again, our ansatz for the background solution is an AdS$_{d+1}$ metric written in slicing coordinates, as in eq. \eqref{SlicingMetric}. 
The Israel junction condition then reads
\begin{equation}\label{IJCDGPf}
    -\frac{d-1}{L}f'(z_b) = -8 \pi G_N \tau + \sigma A  \frac{(d-1)}{L} \left(f(z_b)\right)^2\,,
\end{equation}
where again $\sigma$ denotes the sign of our spacetime \eqref{sigma}.
Notice how the brane position is unaffected in the flat case, where $\sigma=0$, so $\tau=\tau_c$ and the brane remains free. 

\begin{figure}[t]
    \centering
    \includegraphics[scale=0.52]{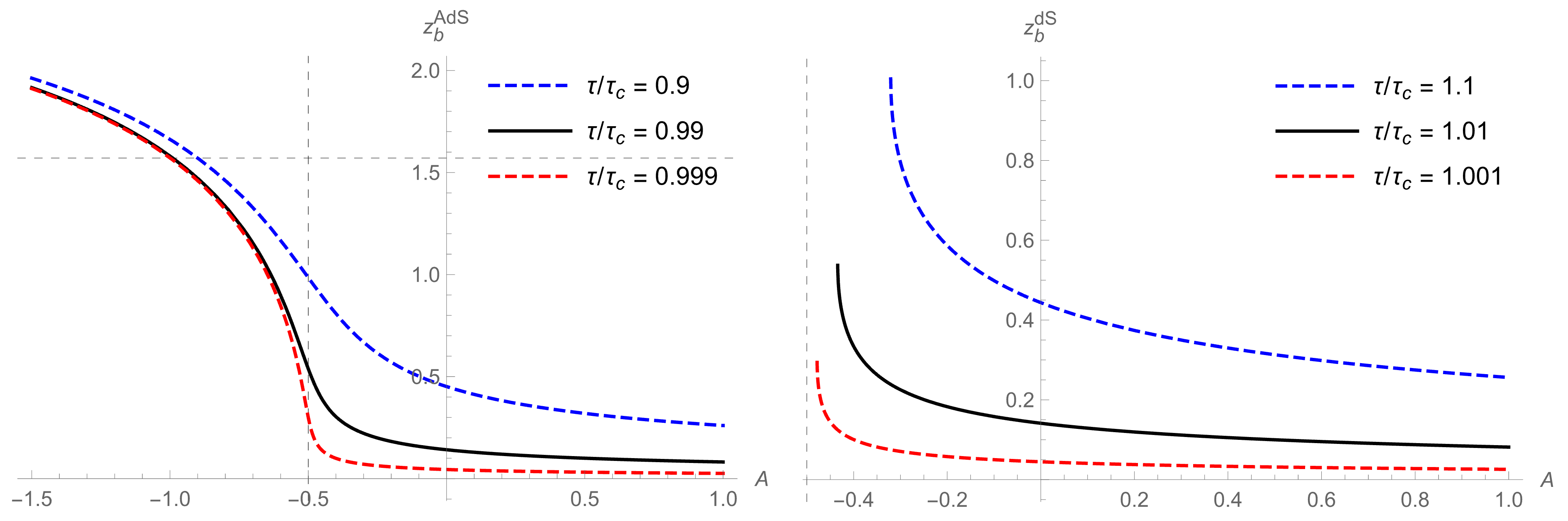}
    \caption{\small{Brane position $z_b$ as a function of $A$, for three different values of $\tau/\tau_c$: $1 + \sigma\tau/\tau_c = 0.1$ (blue dashed), $1 + \sigma\tau/\tau_c = 0.01$ (black), $1 + \sigma\tau/\tau_c = 0.001$ (red dashed). The vertical line corresponds to $A = -1/2$ on both graphs. Positive DGP couplings bring the brane closer to the boundary, while negative DGP couplings are only valid up to some $A \simeq -1/2$. Left: $z_b$ for an AdS brane. Notice how the brane moves to the other side of the spacetime, beyond $z_b = \pi/2$ (horizontal line), at $A \simeq -1/2$. Right: $z_b$ for a dS brane. The brane position has no real solution for $A \lesssim -1/2$.}}
    \label{fig:zBDGPneg}
\end{figure}

For the (A)dS cases, we can relate $f'(z_b)$ and $f(z_b)$ through
\begin{equation}\label{Pyth}
    \left(f(z_b)\right)^2 = \sigma \left( (f'(z_b))^2-1 \right)\,,
\end{equation}
since the functions $f(z)$ are trigonometric. This allows us to solve \eqref{IJCDGPf} for $f'(z_b)$,
\begin{equation}\label{zDGP}
    f'(z_b) = \frac{-1+\sqrt{1+4 A (A + \tau/\tau_c)}}{2A}\,,
\end{equation}
where we choose the plus sign in the quadratic formula because we want $z_b \to 0$ as $\tau \to \tau_c$ for small values of $A$.
Expanding for $A \ll 1$, we obtain
\begin{equation}
    f'(z_b) \simeq \frac{\tau}{\tau_c} + \left(1-\frac{\tau^2}{\tau^2_c} \right) A\, + \mathcal{O}(A^2).
\end{equation}
We see that turning on a positive DGP coupling will bring the brane closer to the boundary, as shown in Fig. \ref{fig:zBDGPneg}.
This behaviour remains true for larger values of $A$, since equation \eqref{zDGP} is monotonous in $A$.
Physically, since the brane has a maximally symmetric geometry, the Ricci scalar $R$ is constant and acts as an extra tension on the brane.

If we choose a negative DGP coupling, however, things change. 
For a small negative DGP coupling, up to approximately $A \sim -1/2$, the brane simply moves slightly away from the boundary.
This behaviour is true for both (A)dS cases, as shown in Fig. \ref{fig:zBDGPneg}.

Now, what happens at larger negatives values of $A$?
For AdS branes with fixed $\tau$, the position of the brane as a function of $A$ is continuous, from $z_b \to \pi$ as $A \to -\infty$, to $z_b \to 0$ as $A \to +\infty$. For values of $\tau$ close to $\tau_c$, the brane remains close to the boundary at $z = 0$ up to some value $A_\text{min} \gtrsim -1/2$, when it rapidly jumps far from $z = 0$. Recall that for AdS branes, $z \in [0,\pi]$, so for $z_b \gtrsim \pi/2$ we are discarding half of our bulk and $d$-dimensional gravity no longer localizes on the brane.

For dS branes, there is a lower limit on $A$ before there is no real solution to \eqref{zDGP},
\begin{equation}
    A_{\text{min}} = -\frac{1}{2}\left(\frac{\tau}{\tau_c} - \sqrt{\frac{\tau^2}{\tau_c^2}-1} \right) \geq -\frac{1}{2}\,.
\end{equation}

Therefore, it makes no sense to consider DGP couplings with $A \lesssim -1/2$ in either case, since then, the position of the brane is either far from $z=0$ or not even well-defined.

%%%%%%%%%%%%%%%%%%%%%%%%%%%%%%%%%%%%%%%%

\subsection{Locally Localized Gravity with DGP}

\begin{figure}[t]
    \centering
    \includegraphics[scale=0.51]{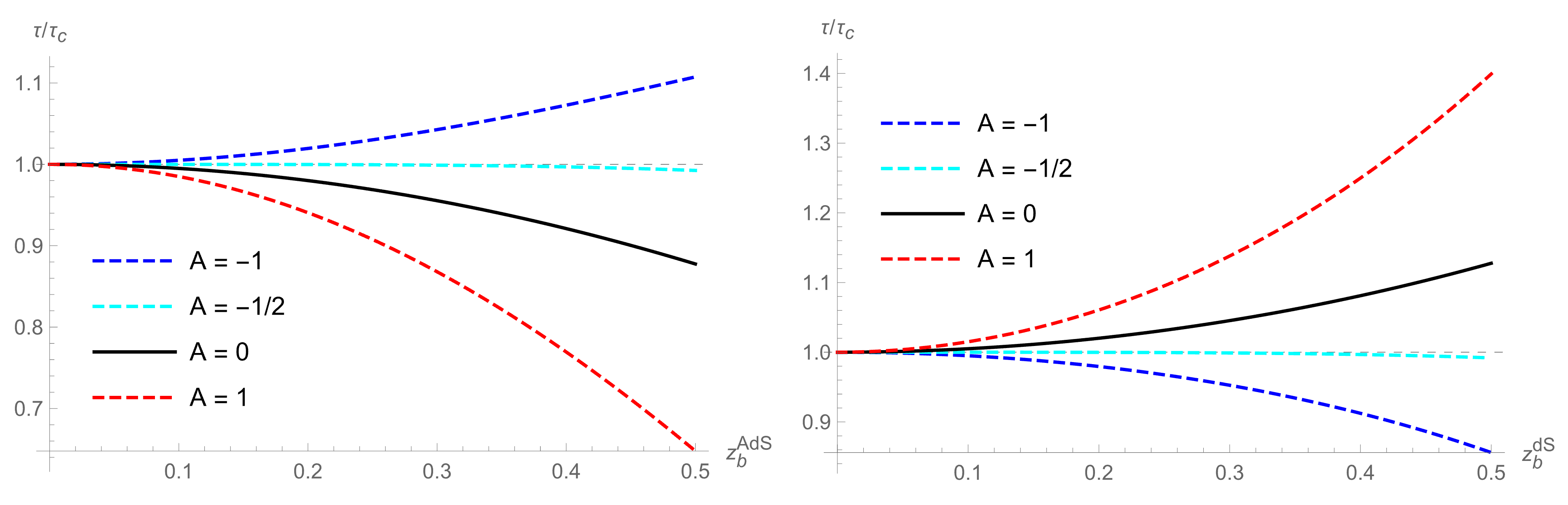}
    \caption{\small{Normalized brane tension $\tau/\tau_c$ as a function of the brane position $z_b$, following eq. \eqref{TauDGPEq}, for four different values of $A$: $A = -1$ (blue dashed), $A = -1/2$ (cyan dashed), $A = 0$ (black), and $A = 1$ (red dashed). The horizontal gray line corresponds to the critical tension $\tau = \tau_c$ on both graphs. Left: $\tau/\tau_c$ for an AdS brane. Notice how we need supercritical tensions to have AdS branes with $A < -1/2$ close to the boundary at $z=0$. Right: $\tau/\tau_c$ for a dS brane. Notice how we now need subcritical tensions to have dS branes with $A < -1/2$ close to the boundary at $z=0$.}}
    \label{fig:TauDGP}
\end{figure}

Previously, we studied the position of the brane $z_b$ as a function of the brane tension $\tau$ and the DGP coupling $A$.
In the following, however, we will work with $z_b$ and $A$ as free parameters, and instead solve \eqref{IJCDGPf} for $\tau$ and tune the brane tension to be
\begin{equation}\label{TauDGPEq}
    \tau = \tau_c \left[ f'(z_b) + \sigma A \left(f(z_b)\right)^2 \right]\,.
\end{equation}
As seen in Figure \ref{fig:TauDGP}, it is now possible to have AdS (dS) branes close to the asymptotic boundary at $z=0$ even with $A \lesssim -1/2$, but only if we allow for supercritical (subcritical) brane tensions.
Nevertheless, by studying their linearized spectrum, we will see that these branes will also show pathological behaviour.

Let us now perturb the bulk metric with axial transverse and traceless perturbations, as in \eqref{PertSlicingMetric}.
Again, we assume $\delta \hat{g}_{ij}(x,z) = H(z) h_{ij}(x)$, with separation constant $E^2$.
Since the bulk Einstein equations are unchanged, we obviously obtain eqs. \eqref{EqBrane} and \eqref{EqRad} again, respectively, for $h_{ij}(x)$ and $H(z)$.
But the Israel junction condition \eqref{IJCDGP} now reads
\begin{equation}
    H'(z_b)h_{ij}(x) + \frac{2 A f(z_b)}{d-2} H(z_b) \left( \hat\Box + 2\sigma \right) h_{ij}(x) = 0\,.
\end{equation}
Luckily we can factor out $h_{ij}(x)$ using the brane equation \eqref{EqBrane} to trade the operator $(\hat\Box + 2\sigma)$ for  the eigenvalue $E^2$. Then, the boundary condition for the radial equation becomes
\begin{equation}\label{BCBraneDGP}
    H'(z_b) + \frac{2 A f(z_b)}{d-2} E^2 H(z_b) = 0\,.
\end{equation}
How will the spectrum of eigenvalues $E^2$ change after this change of boundary conditions?

%%%%%%%

The spectrum will not change much for flat and dS branes.
First, notice that the massless mode remains unchanged in both cases, since we recover the boundary condition $H'(z_b) = 0$ for $E^2 = 0$.
Then, following the volcano potential argument in subsection \ref{sec:EqRad}, one can argue that the continuum of excited eigenvalues will qualitatively have the same properties as if the DGP term were not there. 
Moreover, the solutions to \eqref{EqRad} are still a linear combination of Bessel functions \eqref{HBessel} for the case of flat branes, and a linear combination of hypergeometric functions \eqref{HHyper} for dS branes.
Upon imposing the new boundary condition \eqref{BCBraneDGP}, the only thing that will change is the ratio between the two constants $c_1/c_2$ appearing in the solutions, which will now also depend on the DGP coupling $A$.

%%%%%%%

\begin{figure}[t]
    \centering
    \includegraphics[scale=0.65]{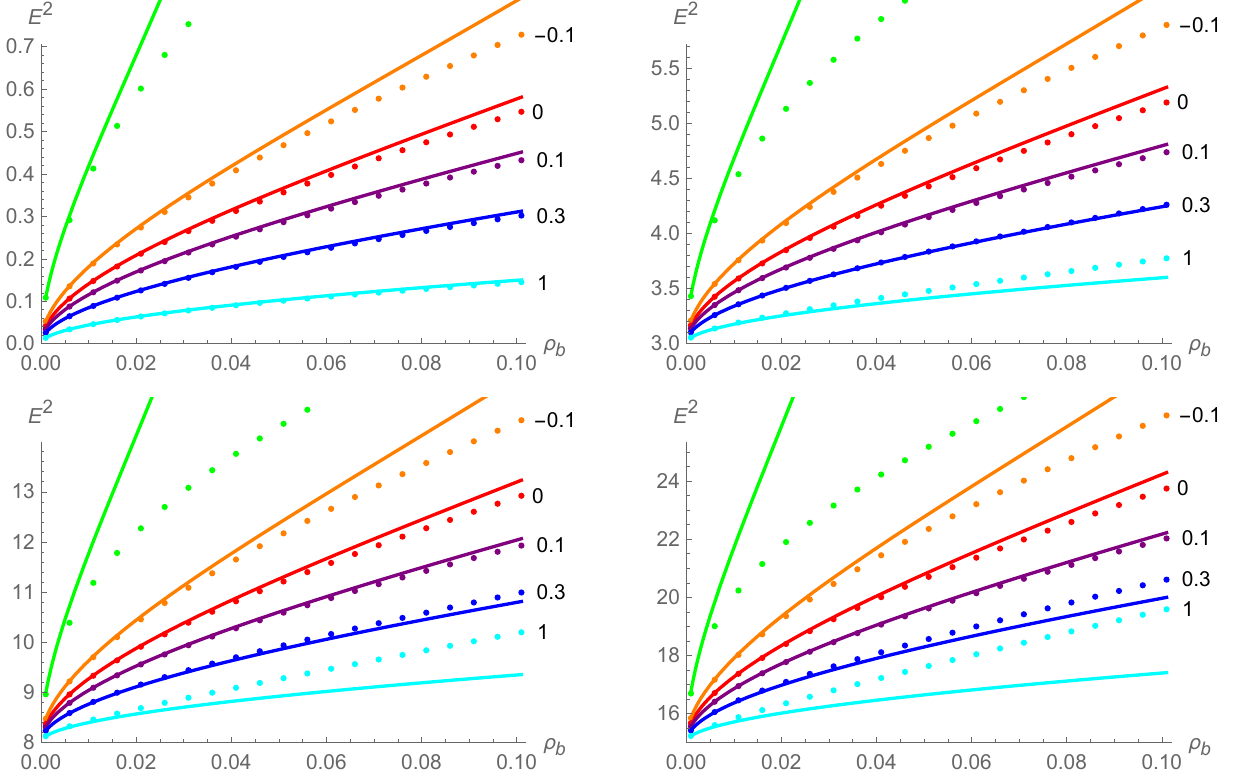}
    \caption{\small{Lowest-lying $d=3$ eigenvalues $E^2_n$ as a function of the brane position $\rho_b$, for different $A$ values. Dots show numerical data, while lines correspond to the improved formula in eq. \eqref{En3DGP}. Green is $A = -0.3$, while the value of $A$ for the other colours is next to the corresponding data.
    Top-left: almost-massless mode, $n = 0$. Top-right: $n = 1$. Bottom-left: $n = 2$. Bottom-right: $n = 3$.
    }}
    \label{EigenDGP}
\end{figure}

On the other hand, for AdS branes, the discrete spectrum of eigenvalues $E^2_{(n,d)}$ changes. 
The solution to equation \eqref{EqRad} for AdS branes must still be \eqref{HLegendrec1c2}, and we are still imposing Dirichlet boundary conditions on the asymptotic boundary, $H(z=\pi) = 0$, which fix the ratio $c_1/c_2$ as shown in \eqref{c2c1}. 
Further imposing the boundary condition \eqref{BCBraneDGP} on the brane discretizes the spectrum, but this discretization now depends on the DGP coupling.

Analytically, in the limit $\rho_b \to 0$, and assuming that $A \ll 1$, we find\footnote{We were also able to find an improved expression for the case of $d=3$ DGP branes,
\begin{equation}\label{En3DGP}
    E^2_{(n,d=3)}(A) \simeq \frac{n(n+2)(1+2A)\pi + (n^2+2n+4)\sqrt{\rho_B}}{(1+2A)\pi - 3\sqrt{\rho_B}}\,.
\end{equation}}
\begin{equation}\label{EndDGP}
    E^2_{(n,d)}(A) \simeq n(n+d-1) + \frac{1}{2}(d-2)(2n+d-1) \frac{\Gamma(n+d-1)}{(\Gamma(d/2))^2\Gamma(n+1)} \frac{\rho_b^{d/2-1}}{(1+2A)}\,.
\end{equation}
Compared to our DGP-free result in eq. \eqref{E2nd}, the only change is the $(1+2A)$ factor on the denominator of the $\rho_b^{d/2-1}$ term.
As shown in Fig. \ref{EigenDGP}, we see that for positive values of $A$, the $\rho_b^{d/2-1}$ term becomes smaller, so $E^2_{(n,d)}$ moves closer to $n(n+d-1)$, while we get the opposite effect for negative values of $A$ up to $A \sim -1/2$, when the expression diverges. 

\begin{figure}[t]
    \centering
    \includegraphics[scale=0.70]{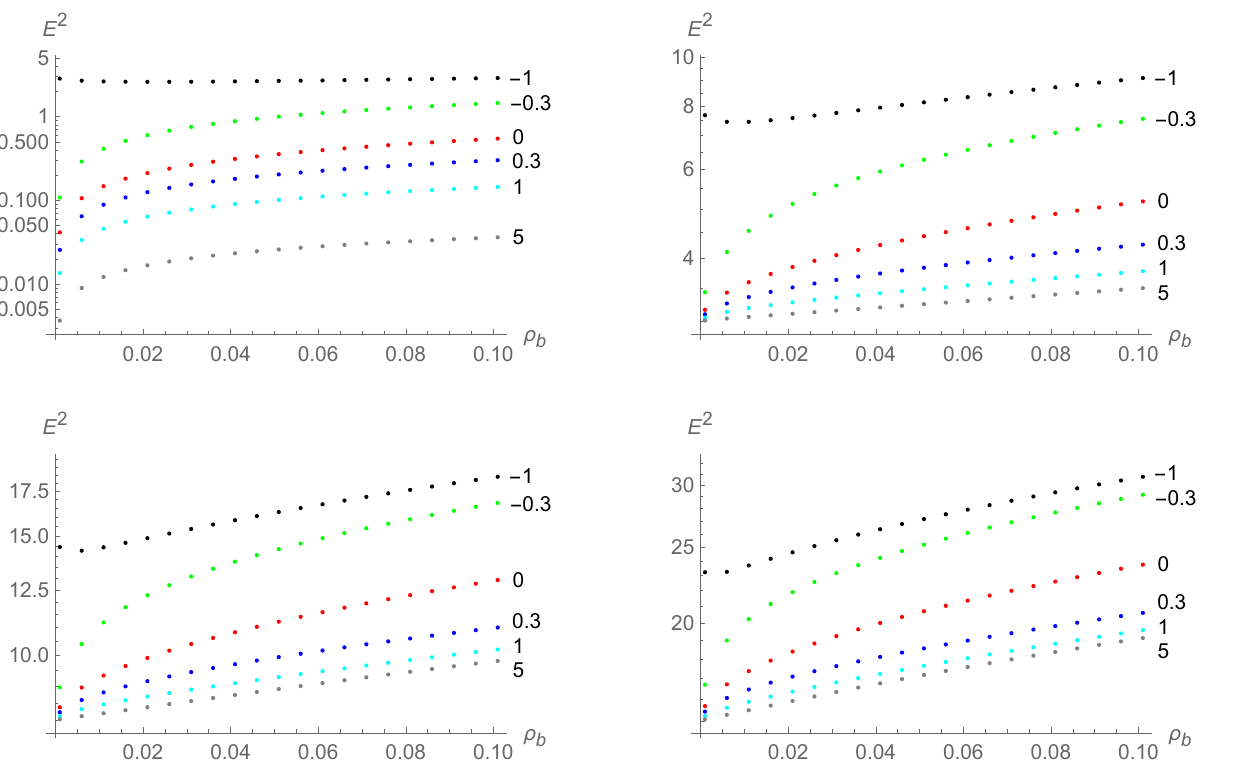}
    \caption{\small{Numerical results in log-scale of the lowest-lying eigenvalues $E^2_n$ for $d=3$ as a function of the brane position $\rho_b$, for different values of $A$, shown next to the corresponding data. Notice how for $A \lesssim -1/2$ the value of the $n$-th eigenvalue jumps from being close to $n(n+2)$ to being closer to the next one, $(n+1)(n+3)$, from below. Top-left: almost-massless mode, with $n = 0$. Top-right: $n = 1$. Bottom-left: $n = 2$. Bottom-right: $n = 3$.
    }}
    \label{EigenLog}
\end{figure}

Numerically, we observe this same behaviour: for positive $A$, the eigenvalues $E^2_{(n,d)}$ move closer to $n(n+d-1)$, while we get the opposite effect for small, negative $A$, with $E^2_{(n,d)}$ moving away from $n(n+d-1)$.
For $A \lesssim -1/2$, the approximation shown in equations \eqref{EndDGP} and \eqref{En3DGP} is no longer valid.
Instead, the eigenvalues $E^2_{(n,d)}$ jump from being close to $n(n+d-1)$ to approaching the next level, $(n+1)(n+d)$, from below, as shown in Fig. \ref{EigenLog}.

In conclusion, turning on a positive DGP coupling, the eigenvalues $E^2_{(n,d)}$ become closer to the ones of empty global AdS.
One can also check that the almost-zero mode becomes more strongly localized on the brane as we turn up $A$, while the opposite is true for the higher overtones.
Turning on a small, negative DGP coupling has a small effect on the eigenvalues, moving them a bit away from the ones of empty global AdS.
Therefore, positive or small enough negative DGP couplings are allowed, and we still obtain $d$-dimensional gravity localized on the brane.
However, if the negative DGP coupling is large enough, $A \lesssim -1/2$, the $E^2_{(n,d)}$ eigenvalues jump away from their value close to $n(n+d-1)$ to the next level.
In particular, this means that we lose the almost-massless mode.
Therefore, gravity no longer localizes on the brane for large negative DGP couplings.

\begin{figure}[t]
    \centering    \includegraphics[scale=0.53]{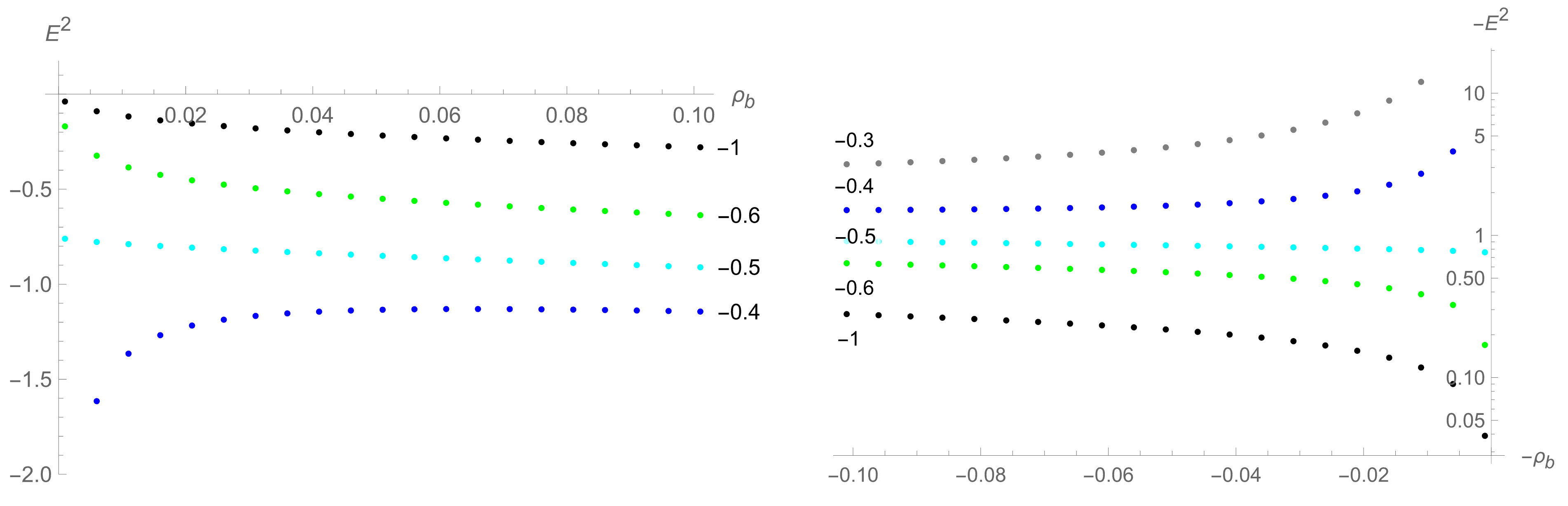}
    \caption{\small{Tachyon mode mass in $d=3$, computed numerically, as a function of $\rho_b$ for different negative values of $A$, labelled next to the respective data.
    For $A \gtrsim -1/2$, the (negative) mass of this mode diverges as $\rho_b \to 0$, which means that this tachyonic mode is beyond the EFT scale of the system, not compromising its validity. 
    For $A \lesssim -1/2$, however, the (negative) mass of this tachyonic mode becomes almost massless as $\rho_b \to 0$.
    Left: Linear plot. Right: Log scale on the vertical axis. We have flipped the signs of both axes to avoid problems with the logarithm.}}
    \label{fig:TachDGP}
\end{figure}

Moreover, numerically and for negative $A$, we have found a negative-mass mode $E^2_t < 0$ in the spectrum, as shown in Fig. \ref{fig:TachDGP}.
For small negative $A$, this eigenvalue has a large (negative) mass, of order $\mathcal{O}(z_b^{-1})$.
From an EFT perspective, it is way beyond the scale at which one should trust the theory, so we believe that this does not compromise the validity of the model.
But again, for $A \lesssim -1/2$, the mass of this tachyonic mode becomes of $\mathcal{O}(1)$, signalling at an instability of the theory, which agrees with all our previous discussions. 

%%%%%%%%%%%%%%%%%%%%%%%%%%%%%%%%%%%%%

\subsection{Induced Gravity on the Brane with DGP}\label{sec:BranePOVDGP}

Following the procedure described in \ref{sec:BranePOV}, the brane effective action now reads
\begin{equation}
    I_{\text{eff}} = \frac{1}{16 \pi G_{N,\text{eff}} } \int d^dx \sqrt{-\gamma} \left[-2\Lambda_{\text{eff}} + \left(1+2A  \right)R + \cdots \right] + I^{UV}_\text{CFT},
\end{equation}
where, $G_{N,\text{eff}}$ and $\Lambda_{\text{eff}}$ are defined in \eqref{GdLd}.
The only change with respect to the effective action written in eq. \eqref{Ieff} is the coefficient in front of the Einstein-Hilbert term.

It is interesting to see how for $A < -1/2$ the Einstein-Hilbert term picks up a minus sign, which again signals to an instability of the theory, since, upon linearization, the graviton picks up the wrong kinetic sign and becomes a ghost. 
This once more agrees with our discussion in previous sections, where we saw that 
the model is not well defined
for $A < -1/2$. 

%%%%%%%%%%%%%%%%%%%%%%%%%%%%%%%%%%%%%%%
%
%
%
%%%%%%%%%%%%%%%%%%%%%%%%%%%%%%%%%%%%%%%

\section{Sending the Brane to the Boundary}

We would now like to make contact with models of dynamical-boundary holography \cite{Compere:2008us, Ishibashi:2023luz, Ishibashi:2023psu, Ishibashi:2024fnm, Ghosh:2023gvc, Ecker:2021cvz, Ecker:2023uea}.
These models need counterterms to render the bulk on-shell action finite, so their action reads
\begin{align}\label{IshiAct}
    I_\text{dyn-bdy} = & \ \frac{1}{16 \pi G_N} \left( \int_\mathcal{M} \df^{d+1} x \sqrt{-G} \left[ R[G] - 2\Lambda^{(d+1)} \right] + 2 \int_{\partial \mathcal{M}_b} d^dx \sqrt{-g} \ K \right) \nonumber \\
    & - \frac{1}{16 \pi G_N} \int_{\partial \mathcal{M}_b} \df^dx \sqrt{-g} \left[ 2 \frac{d-1}{L} + \frac{L}{d-2}R + \cdots \right] \nonumber \\
    & + \frac{1}{16 \pi G^{(d)}_N} \int_{\partial \mathcal{M}} \df^dx \sqrt{-\mathcal{G}} \left( \mathcal{R} - 2\Lambda^{(d)} \right)\,,
\end{align}
where again $G$ denotes the bulk metric, $g$ denotes the induced metric on some hypersurface $\Sigma_{z_b}$ at a finite distance $z_b$ which we will take to the boundary at $z=0$, and $\mathcal{G}$ denotes the boundary metric. 
Notice that we have only written the necessary counterterms up to $d=3$.
Keep in mind that, if our bulk metric is
\begin{equation}
    ds^2_{d+1} = G_{\mu\nu}dy^\mu dy^\nu = \frac{L^2}{\left(f(z)\right)^2} \left[dz^2 + \hat{g}_{ij} dx^i dx^j \right]\,,
\end{equation}
then, since $f(z) \sim z$ near the boundary,
\begin{equation}
    \mathcal{G}_{ij} = \lim_{z \to 0} \hat{g}_{ij} = \lim_{z \to 0} \frac{\left(f(z)\right)^2}{L^2} g_{ij} \simeq \lim_{z \to 0} \frac{z^2}{L^2} g_{ij} \,.
\end{equation}

In their models, the first two lines in the action ---the bulk plus the the usual counterterms--- give rise to a holographic CFT at the boundary, which couples to the dynamical gravity at the boundary given by the third line in the action \eqref{IshiAct}.

%%%%%%%%%%%

\subsection{Adding Counterterms}

%%%%%%%%%%%

In our previous set-up, we had a brane with tension $\tau$ and DGP term $A$ at some finite distance $z_b$. We want to study the models of dynamical-boundary holography by sending the brane to the boundary, but we now need to include counterterms, so our set-up becomes
\begin{align}\label{BrtoBdAct}
    I = & \ \frac{1}{16 \pi G_N} \left( \int_\mathcal{M} \df^{d+1} x \sqrt{-G} \left[ R[G] - 2\Lambda^{(d+1)} \right] + 2 \int_{\partial \mathcal{M}_b} d^dx \sqrt{-g} \ K \right) \nonumber \\
    & - \frac{1}{16 \pi G_N} \int_{\partial \mathcal{M}_b} \df^dx \sqrt{-g} \left[ 2 \frac{d-1}{L} + \frac{L}{d-2}R + \cdots \right] \nonumber \\
    & + \int_{\partial \mathcal{M}_b} \df^dx \sqrt{-g} \left( - \tau + A \frac{L}{8 \pi G_N (d-2)} R \right)\,.
\end{align}
Regrouping terms, and recalling the definition of $\tau_c$ in \eqref{IJCTau}, we can rewrite this as 
\begin{equation}\label{BrToBdAct2}
    I = \Bigg( \text{Bulk + GHY} \Bigg)
    + \int_{\partial \mathcal{M}_b} \df^dx \sqrt{-g }\left[- \left( \tau+\tau_c \right) + \left( A - \frac{1}{2} \right) \frac{L}{8 \pi G_N (d-2)} R + \cdots \right]\,,
\end{equation}
which is the same as what we had in \eqref{ActionBulk} and \eqref{IbrDGP}, but now with tension $\tau \to \tilde{\tau} := \tau+\tau_c$, and DGP coupling $A \to \tilde{A} := A - 1/2$\,.
Therefore, if we include counterterms to our brane action, the sign of the additional brane tension $\tau$ will determine the geometry on the brane, \ie $\tau<0$ will correspond to AdS branes, since then $\tilde\tau < \tau_c$, while $\tau>0$ will correspond to dS branes.
Moreover, we learned that we must have $\tilde{A} \lesssim -1/2$ in order to have a well defined KRS-DGP set-up, and so we need $A>0$.

From a brane-world holographic perspective, since the counterterms exactly cancel the (divergent terms of the) induced brane gravity, the effective action on the brane now reads
\begin{align}\label{BrToBdActEff}
    I_{\text{eff}} & = \int \df^dx \sqrt{-g} \left[ - \tau + A \frac{L}{8 \pi G_{N}(d-2)} R + \cdots \right] + I^{UV}_\text{CFT} \nonumber \\ 
    & = \frac{1}{16 \pi \tilde{G}_{N}} \int \df^dx \sqrt{-g} \left[ - 2 \tilde{\Lambda} + R + \cdots \right]  + I^{UV}_\text{CFT}\,,
\end{align}
where we have defined\footnote{
The fact that $A$ and $\tau$ in \eqref{DefTilde} have no tilde is not a typo. Since, from the bulk perspective, the brane action has a tension $\tilde\tau = \tau + \tau_c$, the $\tau_c$ term precisely cancels the bulk contribution to the brane effective action, and only $\tau$ remains. Similarly, for the DGP coupling $\tilde{A} = A - 1/2$, the $-1/2$ exactly cancels the $R$ contribution to the brane effective action coming from the bulk, and so only $A$ remains.
}
\begin{equation}\label{DefTilde}
    \tilde{G}_N = \frac{(d-2)}{2L}\frac{G_{N}}{A}\,, \quad \quad \quad \tilde{\Lambda} = 8 \pi \tilde{G}_N \tau\,.
\end{equation}
So again we see that the sign of $\tau$ controls the cosmological constant on the brane, and that the value of the DGP coupling $A$ controls the Newton's constant on the brane. 
Again, we cannot have $A < 0$, since then, the Einstein-Hilbert term picks up a minus sign.
These results agree with our analysis in Section \ref{chp:BWsWithDGP}, which now applies to $\tilde\tau$ and $\tilde{A}$. 

\subsection{Brane to Boundary Limit}

Now, we want to send $z_b \to 0$ while keeping the brane dynamical.
Since $g_{ij}$ scales as $L^2/(f(z_b))^2 \sim z_b^{-2}$, the volume element $\sqrt{-g}$ blows up as $L^d/(f(z_b))^d \sim z_b^{-d}$, so we need to rescale $\tau$ and $A$ as we take the limit. Ignoring finite higher-order terms which will go to zero as $z_b \to 0$ for $d=3$,
\begin{align}
    I_{\text{eff}} & = \frac{1}{16 \pi \tilde{G}_N} \int \df^d x \sqrt{-g} \left[ -2 \tilde\Lambda + R + \cdots \right]  + I^{UV}_\text{CFT} \nonumber \\
    & = \frac{1}{16 \pi \tilde{G}_N} \int \df^d x \sqrt{-\hat{g}} \left[ -2 \tilde\Lambda \left( \frac{L}{f(z_b)} \right)^{d} + \left( \frac{L}{f(z_b)} \right)^{d-2} R[\hat{g}] + \cdots \right] + I^{UV}_\text{CFT} \nonumber \\
   & =: \frac{1}{16 \pi \hat{G}_N} \int \df^d x \sqrt{-\hat{g}} \left[ -2 \hat{\Lambda} + R[\hat{g}] + \cdots \right] + I^{UV}_\text{CFT}\,,
\end{align}
where
\begin{equation}
    \hat{G}_N := \left(\frac{f(z_b)}{L} \right)^{d-2} \tilde{G}_N\,, 
    \quad \quad \quad
    \hat\Lambda := \left(\frac{f(z_b)}{L} \right)^{-2} \tilde{\Lambda}\,.
\end{equation}
In terms of the tension and DGP coupling appearing in \eqref{BrtoBdAct}, we see that we need to redefine
\begin{equation}
    A \to \hat A \left(\frac{f(z_b)}{L}\right)^{d-2} \simeq \hat A \left(\frac{z_b}{L}\right)^{d-2}\,, \quad \quad \tau \to \hat \tau \left(\frac{f(z_b)}{L}\right)^{d} \simeq \hat \tau \left(\frac{z_b}{L}\right)^{d}\,.
\end{equation}
so that the action remains finite as we take the limit $z_b \to 0$.

Then, since $\lim_{z_b \to 0} \hat{g}_{ij} = \mathcal{G}_{ij}$, we indeed we get a theory with dynamical boundary gravity coupled to a holographic CFT with no cut-off in the limit where the brane reaches the boundary,
\begin{align}
    I_{\text{eff}} & = \frac{1}{16 \pi \hat{G}_N} \int \df^d x \sqrt{-\hat{g}} \left[ -2 \hat\Lambda + R[\hat{g}] + \cdots \right] + I^{UV}_\text{CFT} \nonumber \\ & \longrightarrow \ \ I_\text{eff-bdy} =  \frac{1}{16 \pi \hat{G}_N} \int \df^d x \sqrt{-\mathcal{G}} \left[ -2 \hat\Lambda + \mathcal{R} \right]  + I_\text{CFT}\,.
\end{align}

Is this well-defined? In the previous section \ref{chp:BWsWithDGP}, we argued that we could not have branes with DGP couplings smaller than $-1/2$, since then either the brane position ceased to be well-defined or the spectrum on AdS branes changed, losing the localization of gravity on the brane.
However, if we want to match with these dynamical-boundary models, we must introduce a DGP term with a coupling of exactly $-1/2$ ---the usual counterterm--- and then add an extra DGP term $A$. 
According to our previous analysis, this should be well-defined if $A$ is positive and large enough so that we are far from this problematic zone of parameters.
However, we have just seen that we need to scale $A$ as $z_b^{d-2}$ to get a finite action as the brane reaches the boundary, so this DGP coupling goes to zero when compared to the counterterm as we take the brane to the boundary.
While this observation is not strong enough to rule out the validity of brane-worlds with a dynamical-boundary, it suggests that they may only work in the strict limit $z_b = 0$, and that the method proposed here is not good enough to interpolate between brane-worlds and dynamical-boundary holography.

%%%%%%%%%%%%%%%%%%%%%%%%%%%%%%%%%%%%%%%%%%%%%%%%%%%%%%%%%%%%%%%%%%%%%%%%%%%%%%%%%%%%%%%%%%%%%
%
%
%
%%%%%%%%%%%%%%%%%%%%%%%%%%%%%%%%%%%%%%%%%%%%%%%%%%%%%%%%%%%%%%%%%%%%%%%%%%%%%%%%%%%%%%%%%%%%%

\section{Adding Higher-Derivative Terms on the Brane}\label{sec:further}

From an EFT perspective, the brane tension and the DGP term are only the first few terms of a series expansion of the brane action.
The next terms would be quadratic in curvature.
So let us now consider a Karch-Randall-Sundrum model with bulk action \eqref{IBulk}, and brane action
\begin{equation}
    I_{\text{brane}} = \int_{\partial\mathcal{M}_b} \df^d x \sqrt{-g} \left[ -\tau + \frac{1}{8 \pi G_N} \left( A \frac{L}{d-2} R + \beta_1 R^2 + \beta_2 R_{ab}R^{ab} + \beta_3 R_{abcd}R^{abcd} \right) \right]\,.
    \label{Ibrbeyond}
\end{equation}
Varying it, we see that the Israel junction condition on the brane now reads
\begin{equation}\label{IJCbeyond}
    K_{ab} - K g_{ab} = - 8 \pi G_N \tau g_{ab} + A\frac{L}{d-2} \left( R g_{ab} - 2R_{ab} \right) + \beta_1 E_{ab}^{(1)} + \beta_2 E_{ab}^{(2)} + \beta_3 E_{ab}^{(3)}\,,
\end{equation}
where the $E_{ab}^{(k)}$ are simply the EOMs of these curvature-squared terms,
\begin{align}
   E_{ab}^{(1)} & = g_{ab}R^2 - 4 R_{ab} R + 4 \nabla_a \nabla_b R - 4 g_{ab} \Box R \,, \\
   E_{ab}^{(2)} & = g_{ab}R_{cd}R^{cd} - 4 R^{cd} R_{acbd} + 2 \nabla_a \nabla_b R - 2 \Box R_{ab} - g_{ab} \Box R \,, \\
   E_{ab}^{(3)} & = g_{ab} R_{cdef}R^{cdef} - 4 R_{a}{}^{cde}R_{bcde} - 8 R^{cd}R_{acbd} + 8 R_{a}{}^c R_{bc}  + 4 \nabla_a \nabla_b R - 4 \Box R_{ab}\,.
\end{align}
Considering an ansatz \eqref{SlicingMetric} with a maximally symmetric brane, the junction condition is
\begin{equation}
    f'(z_b) = \frac{8 \pi G_N L}{d-1} \tau - \sigma A \left(f(z_b)\right)^2
     + \left[ d(d-1)\beta_1 + (d-1)\beta_2 + 2 \beta_3 \right] \frac{(d-4)}{L^3} \left(f(z_b)\right)^4  \,.
\end{equation}
Through the Pythagorean identity \eqref{Pyth}, this becomes a fourth-order equation for $f'(z_b)$. 

One might be worried that upon linearization, due to the complicated appearance of the $E^{(k)}_{ab}$, we cannot longer factor out $h_{ij}(x)$ from the boundary condition on the brane.
However, since we are considering axial transverse traceless perturbations, the linearized junction condition can be completely written in terms of $\Box^n h_{ij}(x)$, $n = 0,1,2$.
Therefore, we can still use the brane equation \eqref{EqBrane} to factor out $h_{ij}(x)$ and get a boundary condition for the radial equation that only depends on $H(z_b)$ and its first derivative,
\begin{equation}
    H'(z_b) + \frac{2 A f(z_b)}{d-2} E^2 H(z_b) + \left[ C_1(E^2) + C_2(E^2) + C_3(E^2) \right] \frac{f(z_b)^3}{L^3} H(z_b) = 0\,,
\end{equation}
where
\begin{align}
    C_1(E^2) & = d(d-1)\beta_1 \left(4+d(d-5)-E^2\right)\,,\\
    C_2(E^2) & = \beta_2 \left((d-4)(d-1)^2-2(d-1)-2E^2+E^4 \right)\,,\\
    C_3(E^2) & = 2 \beta_3  \left((d-1)(d-4) + 2(d-4)E^2 + 2E^4\right) \,.
\end{align}
We postpone the exploration of the allowed values of these couplings for future work ---the important conclusion from this chapter is that it is indeed possible to study the range of validity of higher-derivative operators on the brane action.
Even if the Israel junction condition will get higher-derivative terms, it will always be possible to factor out $h_{ij}(x)$ using \eqref{EqBrane} to get a boundary condition for the radial equation in terms only of $H(z_b)$ and $H'(z_b)$.

%%%%%%%%%%%%%%%%%%%%%%%%%%%%%%%%%%%%%%%%%%%%%%%%%%%%%%%%%%%%%%%%%%%%%%%%%%%%%%%%%%%%%%%%%%%%%
%
%
%
%
%
%
%
%
%
%%%%%%%%%%%%%%%%%%%%%%%%%%%%%%%%%%%%%%%%%%%%%%%%%%%%%%%%%%%%%%%%%%%%%%%%%%%%%%%%%%%%%%%%%%%%

\section{Conclusions and Future Directions}

In this work, we have generalized and expanded on results on $d\geq3$ brane-worlds with all three possible maximally symmetric brane geometries, both from the bulk perspective and the dual brane perspective. 

We have investigated how including a DGP term affects the localization of gravity on the brane, searching for pathologies in the theory. 
This has allowed us to put a bound on the DGP coupling.
A positive coupling is always permitted, and sufficiently small negative couplings are allowed too.
Nevertheless, the model is unstable for large negative DGP couplings.
If we keep the brane tension fixed while changing the DGP coupling $A$, we have seen that, if $A \lesssim -1/2$, the position of the brane is not well-defined.
For AdS branes, it jumps to the other side of the bulk, close to $z = \pi$, while
there cannot be dS branes with $A < -1/2$, since there is no real solution for the position of the brane in the bulk.
If instead we keep the position of the brane fixed while changing $A$, we have found that, if $A < -1/2$, then gravity no longer localizes on AdS branes.
The almost-zero mode is lost, and instead we get a tachyon in the spectrum with a negative small mass of $\mathcal{O}(1)$.
These results are consistent with the effective action of the dual brane picture, in which the Einstein-Hilbert term picks up the wrong sign if $A < -1/2$.

We have also explored whether one can use these DGP-KRS brane-world models to study dynamical-boundary holography \cite{Compere:2008us, Ishibashi:2023psu, Ishibashi:2024fnm, Ishibashi:2023luz, Ghosh:2023gvc, Ecker:2021cvz, Ecker:2023uea},
by sending the brane to the boundary while rescaling its tension and DGP coupling so that these operators remain finite at infinity.
Now, however, since we are reaching the boundary, we need to add counterterms so that $I_\text{CFT}$ does not diverge.
Usually, counterterms are added on a regulating hypersurface at some finite $\varepsilon$ distance from the boundary, and then the limit $\varepsilon \to 0$ is taken.
It seems natural, then, to add these counterterms on the brane, which is itself a regulating surface at some distance $\rho_b \to 0$.
The problem is that, seen as operators on the brane, the first few counterterms are precisely a critical tension $\tau_c$ and a DGP term with coupling $A = -1/2$, where we saw that the DGP-KRS construction breaks down.
It seems then that DGP-KRS brane-worlds cannot be used as a way to study dynamical-boundary holography. However, this result could also hint at possible hidden pathologies in holographic models with dynamical boundary. Further research is needed in this direction.

%%%%%%%%

Besides studying the brane-to-boundary limit more in-depth, 
this work offers many other interesting extensions.
One possibility would consist in studying brane-world holography with higher order operators on the brane beyond the DGP term, as we started doing at the end of this paper. 
This is also a necessary step if one wants to use these brane-world constructions to study dynamical-boundary holography with $d \geq 4$, since then we already need curvature-squared counterterms.
One could also consider brane-world holographic models in which the bulk theory is not Einstein gravity but a higher-curvature theory of gravity, while a different route would be considering alternative boundary conditions on the brane.

%%%%%%%%%%%%%%%%%%%%%%%%%%%%%%%%%%%%%%%%%%%%
%
%%%%%%%%%%%%%%%%%%%%%%%%%%%%%%%%%%%%%%%%%%%%

\subsection*{Acknowledgements}

The author thanks D. Neuenfeld and especially R. Emparan ---who proposed the research topic--- for useful discussions and comments on the draft.
This work was partly supported by the Spanish Ministry of Universities through FPU grant No. FPU19/04859.

%%%%%%%%%%%%%%%%%%%%%%%%%%%%%%%%%%%%%%%%%%%%%%%%%%%%%
%
%%%%%%%%%%%%%%%%%%%%%%%%%%%%%%%%%%%%%%%%%%%%%%%%%%%%%

\bibliography{Bib.bib}

%%%

\end{document}